\pdfoutput=1
\documentclass[
    aps,pre,twocolumn,
    reprint,
    superscriptaddress,
    floatfix,
    amssymb,
    longbibliography
]{revtex4-2}

\usepackage{amsmath}
\usepackage{amsfonts}
\usepackage{graphicx}
\usepackage{bm}

\usepackage[utf8]{inputenc}

\usepackage{nicefrac}

\usepackage[normalem]{ulem}

\usepackage{xcolor}

\definecolor{linkColor}{rgb}{0,0.3,0.7}
\usepackage{hyperref}
\hypersetup{colorlinks=true,
            allcolors=linkColor,
            pdfborder={0 0 0},
            pdfencoding = auto
            }

\def\etaStat{\eta_\mathrm{stat}}

\begin{document}

\title{Chemotaxis-induced phase separation}

\author{Henrik Weyer}
\thanks{H.W.\ and D.M.\ contributed equally to this work.}
\author{David Muramatsu}
\thanks{H.W.\ and D.M.\ contributed equally to this work.}
\affiliation{Arnold Sommerfeld Center for Theoretical Physics and Center for NanoScience, Department of Physics, Ludwig-Maximilians-Universit\"at M\"unchen, Theresienstra\ss e 37, D-80333 M\"unchen, Germany}
\author{Erwin Frey}
\email{frey@lmu.de}
\affiliation{Arnold Sommerfeld Center for Theoretical Physics and Center for NanoScience, Department of Physics, Ludwig-Maximilians-Universit\"at M\"unchen, Theresienstra\ss e 37, D-80333 M\"unchen, Germany}
\affiliation{Max Planck School Matter to Life, Hofgartenstraße 8, D-80539 Munich, Germany}

\date{September 10, 2025}
	
\begin{abstract}
    Chemotaxis allows single cells to self-organize at the population level, as classically described by Keller--Segel models.
    We show that chemotactic aggregation can be understood using a generalized Maxwell construction based on the balance of density fluxes and reactive turnover.
    This formulation implies that aggregates generically undergo coarsening, which is interrupted and reversed by cell growth and death.
    Together, both stable and spatiotemporally dynamic aggregates emerge.
    Our theory mechanistically links chemotactic self-organization to phase separation and reaction--diffusion patterns.
\end{abstract}

\maketitle

Chemotaxis is a prevalent strategy in biology that enables both  microbes and tissue cells to move collectively and directionally in response to chemical gradients~\cite{Berg2004,Murray2003,Gerard.Rollins2001,Roussos.etal2011,Trepat.etal2012}.
This can lead to self-organized aggregation when cells respond to gradients of self-secreted signaling molecules, described by the foundational Keller--Segel (KS) model \cite{Keller.Segel1970}.
Diverse model variants account for different chemotactic responses, crowding effects, and cell proliferation \cite{Hillen.Painter2009}.
Without cell proliferation and death the chemotactic aggregates are observed to undergo a coarsening process \cite{Hillen.Painter2009,Kang.etal2007,Kavanagh2014,Meyer.etal2014,Rapp.Zimmermann2019,OByrne.Tailleur2020}.
In contrast, cell growth and death induce a fixed pattern wavelength leading to intricate patterning \cite{Mimura.Tsujikawa1996,Hillen.Painter2009} and spatiotemporal chaos \cite{Painter.Hillen2011,Ei.etal2014}, as well as an absorbing-state phase transition with distinct scaling regimes \cite{vanderKolk.etal2023}.
However, a unifying understanding of chemotactic aggregates and their long-time dynamics in general KS models is missing.
It remains unknown whether a generic mechanism links coarsening, the selection of fixed intrinsic wavelengths, and the spatiotemporally dynamic patterns.

In this letter, we derive the generic behavior of chemotactic aggregates in KS models by developing a combined description of the counteracting effects of random (diffusive) and chemotactic (directed) cell movement.
Stationary aggregates are thereby constructed in phase space based on a reactive area balance representing a generalized Maxwell construction.
We then argue that the resulting net cell movement drives the coarsening of collections of well-separated aggregates.
(Slow) cell growth and death interrupt and reverse the coarsening process, resulting in both fully nonlinear steady-state patterns with fixed wavelengths (aggregate sizes) and spatiotemporally dynamic aggregates.
Our theory is independent of the specific mathematical form of the models and relates the macroscopic dynamics of chemotactic aggregates to (active) phase separation \cite{Glotzer.etal1995,Zwicker.etal2015,Li.Cates2020} and (mass-conserving) reaction--diffusion systems \cite{Brauns.etal2021,Weyer.etal2023}.

We consider a chemotactic population $\rho(\mathbf{x}, t)$ subject to diffusion, with diffusion coefficient $D_\rho$, and chemotaxis, modeled as advection along the gradients in the chemoattractant field $c(\mathbf{x}, t)$ \cite{Keller.Segel1970}:
\begin{equation}
\label{eq:KS-rho}
    \partial_t \rho(\mathbf{x},t) = D_\rho \nabla^2 \rho - T\nabla \big[\chi_\rho(\rho)\chi_c(c)\rho\nabla c\big] + \varepsilon s(\rho).
\end{equation}
Here, ${T\chi_\rho(\rho)\chi_c(c) > 0}$ represents the strength of chemotaxis with the positive sensitivity functions $\chi_\rho$ and $\chi_c$.  
For example, $\chi_\rho$ may describe crowding effects that suppress chemotaxis, while $\chi_c$ may capture nonlinear kinetics of the chemotaxis receptors~\cite{Hillen.Painter2009}.
Cell growth and death are described by $s(\rho)$ with a dimensionless strength $\varepsilon$.
The dynamics of the chemoattractant, produced by the cells and subject to degradation, is given by ${\partial_t c(\bm{x},t) = D_c \nabla^2 c + f(\rho,c)}$.
To model net chemoattractant production by the cells, we assume ${\partial_\rho f > 0}$, biasing $f$ toward chemoattractant production with increasing cell density $\rho$.

Random cell motion and chemotaxis can be captured in one quantity by rewriting Eq.~\eqref{eq:KS-rho} as
\begin{equation}\label{eq:cont-eq}
    \partial_t \rho(\mathbf{x},t) = T\nabla \big[\chi_\rho \rho\nabla \eta\big] + \varepsilon s(\rho) \, ,
\end{equation}
introducing a \emph{mass-redistribution potential} (cf.\ Refs.~\cite{Otsuji.etal2007,Brauns.etal2020,Cotton.etal2022}) 
\begin{equation}\label{eq:eta}
    \eta(\rho(\mathbf{x},t),c(\mathbf{x},t)) := 
    \frac{D_\rho}{T}\int_{\rho_0}^{\rho(\mathbf{x},t)}
    \!\!\!\!\!\! \mathrm{d}\rho \, \frac{1}{\chi_\rho \rho}
    -\int_{c_0}^{c(\mathbf{x},t)}  \!\!\!\!\!\! \mathrm{d}c \, \chi_c\, .
\end{equation}
Here, we defined two arbitrary reference densities $\rho_0$ and $c_0$.
The mass-redistribution potential is a function of the local densities $\rho$ and $c$ and thus is a spatially varying field. 
Comparing the sign of the flux term in Eq.~\eqref{eq:cont-eq} with the sign of the chemotaxis term in Eq.~\eqref{eq:KS-rho}, it is evident that $\eta(\mathbf{x}, t)$ acts as an effective chemorepellent.
The effective field $\eta$ integrates the counteracting effects of diffusion and chemoattraction: it increases with $\rho$, promoting spreading from high to low densities, and decreases with $c$, driving aggregation toward regions of high chemoattractant concentration [Eq.~\eqref{eq:eta}].

Mathematically, the mass-redistribution potential $\eta$ resembles the chemical potential in binary phase separation with the mobility $\chi_\rho \, \rho$ [cf.\ Eq.~\eqref{eq:cont-eq} for ${\varepsilon = 0}$ with Model B] \cite{Hohenberg.Halperin1977,Bray2002}. 
However, $\eta$ is not the derivative of a free energy functional but follows its own dynamic equation; see~\cite{supplemental_material}.
For particular model choices, one can find a relation to a mathematical variational structure \cite{Horstmann2001,Wolansky2002}.
The numerical analysis employs the minimal Keller--Segel (mKS) model with ${\chi_\rho = \chi_c=1}$ and a linear chemoattractant production and decay term ${f = \rho-c}$ \cite{Childress.Percus1981}.
Then, one has ${\eta = \frac{D_\rho}{T}\log\rho - c}$.
We start our analysis with the mass-conserving dynamics where ${\varepsilon = 0}$.

\begin{figure}
	\includegraphics{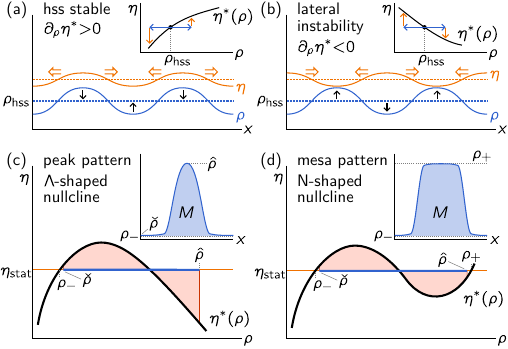}
	\caption{
    The mass-redistribution potential (orange) induces redistribution (orange arrows) of the cell density (blue) that counteracts (a) or amplifies (b) initial perturbations (black arrows).
	(c) $\mathsf{\Lambda}$-shaped NCs $\eta^*(\rho)$ lead to stationary peak patterns.
    The low-density intersection between the NC (black) and the flux-balance subspace ${\eta = \etaStat}$ (orange) determines the low-density plateau $\rho_-$ (pattern minimum ${\check{\rho}\gtrsim \rho_-}$).
    The value $\etaStat$ is determined by the reactive area balance (balance of the red-shaded areas).
    (d) $N$-shaped NCs give an additional high-density NC-FBS intersection $\rho_+$ (pattern maximum ${\hat{\rho}\lesssim \rho_+}$).
    The resulting pattern is mesa-shaped.
	}
	\label{fig:1}
\end{figure}

\textit{Lateral instability.\;---}
A uniform cell density (homogeneous steady state, hss) is unstable if chemotactic drift dominates over diffusive spreading, amplifying small density perturbations.
From Eq.~\eqref{eq:cont-eq}, this requires that $\eta$ decreases where $\rho$ increases, driving net flux toward denser regions [orange arrows in Fig.~\ref{fig:1}(a,b)].
Although ${\eta}$ increases with ${\rho}$ at fixed chemoattractant concentration ${c}$ (diffusive contribution), cell-density perturbations also alter ${c}$ through production, potentially causing ${\eta}$ to decrease where ${\rho}$ increases [chemotactic contribution, second term in Eq.~\eqref{eq:eta}].
For long-wavelength modulations, mass redistribution is slow compared to the local relaxation of ${c}$, which quickly adjusts to a \textit{quasi-steady state} (QSS) ${c^*(\rho)}$ defined by the nullcline ${f(\rho, c^*(\rho)) = 0}$.
Accordingly, ${\eta}$ locally relaxes to ${\eta^*(\rho) := \eta(\rho, c^*(\rho))}$, and chemotactic aggregation becomes self-amplifying if the effective slope is negative:
\begin{equation}\label{eq:slope-crit}
    \partial_\rho \eta^*(\rho_{\mathrm{hss}}) < 0\, .
\end{equation}
This \emph{nullcline-slop criterion} also holds in the short-wavelength regime (see Appendix~\ref{app:slope-criterion}).
The instability resembles the mass-redistribution instability found in mass-conserving reaction--diffusion systems as well as spinodal decomposition~\cite{Brauns.etal2020,Weyer.etal2023}.

\textit{Stationary aggregate profiles.\;---}
The stationary, one-dimensional pattern profile $[\Tilde{\rho}(x), \Tilde{c}(x)]$ can be constructed in the $(\rho,\eta)$-phase plane by generalizing local equilibria theory \cite{Halatek.Frey2018,Brauns.etal2020}.
In steady state, the diffusive and chemotactic fluxes must balance, i.e., the total flux ${\sim \partial_x\eta}$ [cf.\ Eq.\eqref{eq:cont-eq}] must vanish throughout domains with no-flux or periodic boundary conditions.
Thus, stationary patterns are constrained to the \emph{flux-balance subspace} (FBS) ${\Tilde{\eta}(x)=\eta_\mathrm{stat}}$ [Fig.\ref{fig:1}(c,d); $\eta_\mathrm{stat}$ is discussed below].
(Well-separated) aggregates are either mesa-shaped---with low- and high-density plateaus connected by interfaces [Fig.\ref{fig:1}(d)]---or peaks, where density maxima do not saturate in plateaus [Fig.\ref{fig:1}(c)].
Plateaus and inflection points, where the curvature $\partial_x^2 \Tilde{c}$ (approximately) vanishes, are determined by FBS–NC intersections and serve as landmarks of the pattern.
In particular, the outer intersections $\rho_\pm$ of a \textsf{N}-shaped NC are approached exponentially by the pattern extrema $\check\rho$ and $\hat\rho$ as plateau widths increase [sharp-interface approximation in \cite{supplemental_material}; Fig.\ref{fig:1}(d)].
Peak patterns occur if the rightmost FBS–NC intersection $\rho_+$ does not exist (as in a $\mathsf{\Lambda}$-shaped NC) or lies at higher densities (asymmetric \textsf{N}-shaped NC) than the total domain mass $M$ set by initial conditions allows for [blue-shaded in Fig.\ref{fig:1}(c,d)].
For example, the mKS model has a $\mathsf{\Lambda}$-shaped NC allowing unrestricted cell-density growth, while receptor-binding kinetics \cite{Segel1977} or volume-filling effects \cite{Hillen.Painter2001,Painter.Hillen2002} lead to \textsf{N}-shaped NCs, directly explaining their density saturation.
A mathematical criterion for the transition from peak to mesa patterns is given in Ref.~\cite{Hillen2007}.

The reaction NC $\eta^*(\rho)$ separates the $(\rho,\eta)$ phase plane into regions of net chemoattractant production (${f > 0}$) and degradation (${f < 0}$).
Since $\eta$ decreases with $c$ at fixed $\rho$ [Eq.~\eqref{eq:eta}], points above the NC correspond to degradation, and points below to production (multistability is discussed in~\cite{supplemental_material}).
The FBS position $\eta_\mathrm{stat}$---previously introduced as an integration constant~\cite{Schaaf1985}---sets the vertical placement of the pattern in this phase plane, and thus the relative contributions of production and degradation. For the pattern to be stationary, $\eta_\mathrm{stat}$ must self-adjust to balance total production and degradation.
This turnover balance is captured by a Maxwell-like area rule between FBS and NC [Fig.~\ref{fig:1}(c,d); see~\cite{supplemental_material}];
for example, in the mKS model:
${0 = \int_{\check{\rho}}^{\hat{\rho}}\mathrm{d}\rho \, [\eta_\mathrm{stat}-\eta^*(\rho)]/\rho}$.

\begin{figure*}[tb]
	\includegraphics{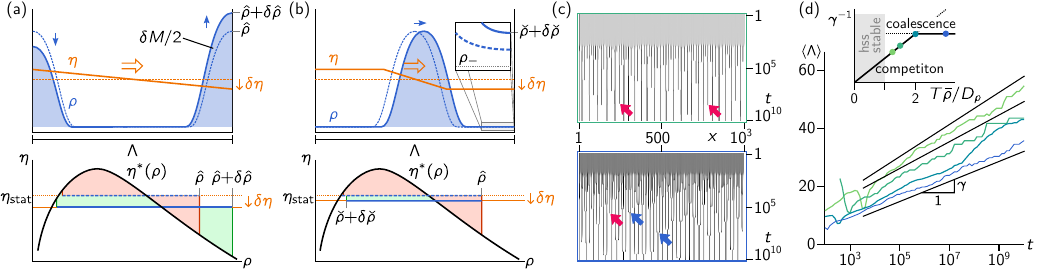}
	\caption{
	Illustration of the mass-competition instability for the competition of two peaks (a) and two plateaus (b; coalescence of the peak with its mirror image) destabilizing the symmetric pattern (dashed lines).
    In both cases (a) and (b), the changes in the stationary mass-redistribution potentials $\delta\eta$ of the two elementary patterns with the peak mass (blue-shaded areas) and plateau lengths lead to self-amplifying mass transport (orange).
    The generalized Maxwell area balance (shown for the right elementary patterns in the lower panels) shows that the (quasi-)stationary mass-redistribution potential decreases at the larger peak and shorter plateau (red- and green-shaded areas).
    (c) In a large system, these instabilities induce a coarsening process of the pattern by competition (examples of peak collapse marked by red arrows) and coalescence (blue arrows) as shown in two kymographs of simulations with $T\Bar{\rho}/D_\rho= 1.5, 3$.
    The faster mass-competition process [cf.\ inset in (d)] dominates the dynamics.
    (d) During the coarsening process, the average pattern wavelength $\langle\Lambda\rangle(t)$ continuously grows.
    Numerical simulations at $T\Bar{\rho}/D_\rho=1.25, 1.5, 2, 3$ (average over 3, 1, 8, and 2 replicates; light green to dark blue) are compared with the prediction ${\langle\Lambda\rangle(t)\sim \max(\gamma_\pm) \log(t)}$ with ${\gamma \propto 5/8,3/4,1,1}$ (black lines; see \cite{supplemental_material}).
    Apart from the average density $\Bar{\rho}$ specified above, the simulation parameters are $T = 5$, $D_\rho = 0.1$, $D_c = 1$, with system length $L= 1000$.
	}
	\label{fig:2}
\end{figure*}

\textit{Coarsening due to mass competition.\;---}
In large systems, the KS systems form multiple (quasi-stationary) aggregates.
Intriguingly, a multitude of numerical and mathematical work on specific models showed that these multiple aggregates are not stable but undergo a coarsening process, paralleling Ostwald ripening and coalescence of droplets in phase-separating mixtures \cite{Meyer.etal2014,OByrne.Tailleur2020,Dinelli.etal2024,Rapp.Zimmermann2019,Dolak.Schmeiser2005,Potapov.Hillen2005,Kang.etal2007,Kavanagh2014}.
Our framework based on the reactive area balance identifies the model-independent mechanism that explains uninterrupted coarsening in mass-conserving KS systems.

Coarsening proceeds via competition between and coalescence of neighboring aggregates, exemplified for peak patterns in Fig.~\ref{fig:2}(a,b).
Our analysis (see Appendix~\ref{app:coarsening-quant}) shows that these processes are caused by \emph{mass competition}: the self-amplifying mass transport along gradients in $\eta$, either between aggregates or from one side of the aggregate toward the other.
Mass competition arises because the reactive area balance generically implies that $\eta_\mathrm{stat}$ decreases both with the mass ${M=\int\mathrm{d}x(\rho-\rho_-)}$ (height) of half-mesas and -peaks (elementary patterns)---at fixed length of the low-density plateau [Fig.~\ref{fig:2}(a)], and also with decreasing length of the low-density plateau---at fixed mass [Fig.~\ref{fig:2}(b)].
Within a QSS approximation, $\eta$ thus decreases during mass competition at the elementary pattern with the already larger peak mass [Fig.~\ref{fig:2}(a)] or shorter low-density plateau [Fig.~\ref{fig:2}(b)], resulting in additional mass transport toward these and an amplification of any initial asymmetry.
We quantify the growth rates ${\sigma_\pm>0}$ of this \emph{mass-competition instability} in the competition and coalescence scenario, respectively, in Appendix~\ref{app:coarsening-quant} and extend the analysis beyond the QSS approximation in Ref.~\cite{Weyer.etalinpreparationa}.
This instability makes uninterrupted coarsening a generic feature of mass-conserving Keller–-Segel systems and allows its quantification [see Fig.~\ref{fig:2}(c,d) and Appendix~\ref{app:coarsening-mKS}].
It unifies previous results for specific systems \cite{Dolak.Schmeiser2005,Potapov.Hillen2005,Kang.etal2007,Kolokolnikov.etal2014,Kong.etal2024,Meyer.etal2014,OByrne.Tailleur2020} and likewise accounts for phase-separation-like dynamics in systems where chemotactic drift arises from quorum-sensing-regulated motility~\cite{Fu.etal2012,Tailleur.Cates2008} or from cross-diffusion couplings~\cite{Cotton.etal2022}.
Thus, our framework establishes chemotaxis-induced phase separation as part of the broader class of coarsening phenomena in nonequilibrium pattern-forming systems.

\begin{figure*}[bt]
	\includegraphics{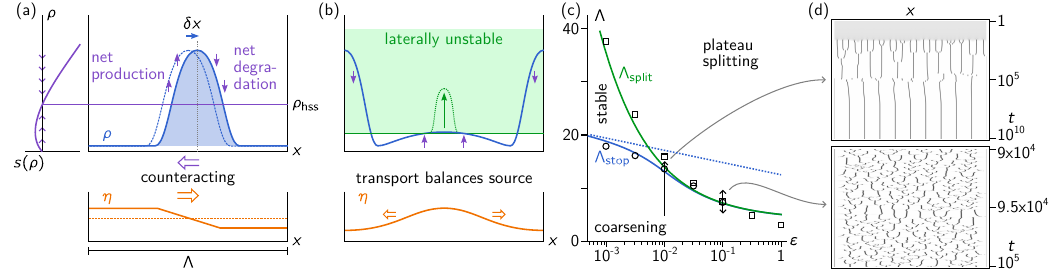}
	\caption{
	(a) Illustration of interrupted coarsening due to production and degradation [$s(\rho)$; purple] counteracting the mass-competition instability (orange) which destabilizes the symmetric peak position against shifts $\delta x$.
    (b) Plateau splitting occurs when the plateau density enters the laterally unstable region (green) due to production in the low-density plateau.
    In mesa patterns, high-density plateaus can split as well due to degradation when the plateau density decreases into the laterally unstable density regime.
    (c) The stability of patterns with wavelength $\Lambda$ in the mKS model with a logistic source term ${s(\rho)=\rho (1-\rho)}$ of strength $\varepsilon$.
    The threshold of plateau splitting (green) lies at low wavelengths $\Lambda$ and deforms the threshold of interrupted coarsening at large source strength $\varepsilon$ (solid compared to dashed blue line).
    The analytic thresholds (lines) well describe the numerical thresholds (circles and squares) obtained by numerical continuation and linear stability analysis (see \cite{supplemental_material}).
    (d) The kymographs show $\log(1+\rho)$ (grayscale, $0$ to $1$) for $\varepsilon = 0.01, 0.1$ simulated on a domain with reflective boundaries and length $L=149, 74$ corresponding (approximately) to a multiple of the mean of the numerical thresholds $\Lambda_\mathrm{stop,split}$.
    The same simulation parameters are used as in Fig.~\ref{fig:2}(d).
	}
	\label{fig:3}
\end{figure*}

\textit{Source terms interrupt coarsening.\;---}
To account for cell growth and death, we now discuss the limit of weakly broken mass conservation (${\varepsilon \ll 1}$).
Here, the average total density $\Bar{\rho}_\mathrm{stat}$ of the stationary pattern is determined by the balance of overall production and degradation, ${\int_\Omega \mathrm{d}x \, s(\tilde{\rho})=0}$ (source balance), not by the initial condition. 
Thus, a pattern only forms if ${0 = s(\rho_\mathrm{hss})}$ holds at a density ${\check{\rho}<\rho_\mathrm{hss}<\hat{\rho}}$.
Otherwise the system will be homogeneously filled or depleted of cells due to excessive cell growth or death.
Moreover, for stability of the elementary (half-mesa/-peak) pattern, the source term $s$ must drive any deviations $\Bar{\rho}-\Bar{\rho}_\mathrm{stat}$ from the average total density that balances production and degradation back to zero. 
We therefore assume net cell growth at low and net death at high densities as, for example, in logistic growth [cf.\ Fig.~\ref{fig:3}(a)].
As mass competition amplifies mass differences, and hence differences in $\Bar{\rho}$ between elementary patterns, the source term counteracts the mass-competition instability.

To illustrate the counteracting effect, consider the coalescence of a peak with its neighbor (mirror image) [Figs.~\ref{fig:2}(b),~\ref{fig:3}(a)], the dominant coarsening pathway at large $\Bar{\rho}_\mathrm{stat}$ in the mKS model [Fig.~\ref{fig:2}(c,d)].
If the peak shifts right by $\delta x$, the right plateau shortens, reducing the total production of cells in the right plateau by approximately ${-\varepsilon s(\rho_-)\delta x}$ while production increases by the same amount to the left of the peak, as follows from integration of Eq.~\eqref{eq:cont-eq}.
As a result, the chemotactic flux of new cells into the aggregate, that balances the degradation of cells inside, is asymmetric and larger on the left side of the aggregate.
Thus, the aggregate will shift back towards the symmetric position because more cells accumulate on its left side [Fig.~\ref{fig:3}(a)].
To this stabilizing, source-induced shift process the destabilizing mass-competition process due to the $\eta$ gradients in the first term of Eq.~\eqref{eq:cont-eq} [Fig.~\ref{fig:2}(b)] must be added.
To lowest order in $\varepsilon$, the competition effect can be described using the result for the mass-conserving system (see Appendix~\ref{app:IC}).
Mass competition is suppressed if the growth rate of the mass-competition instability under the influence of weak source terms fulfills ${\sigma_\varepsilon^-(\Lambda) \leq 0}$.
Because the mass-competition rate ${\sigma^{-}(\Lambda)}$ typically decreases strongly in $\Lambda$ (exponentially in the mKS model), the mass-competition instability is suppressed above a threshold wavelength $\Lambda_\mathrm{stop}$ fulfilling ${\sigma_\varepsilon^-(\Lambda_\mathrm{stop}) = 0}$ [dotted line in Fig.~\ref{fig:3}(c)].
Similar formulas follow from the same reasoning for the competition between peaks and for mesa patterns.

\textit{Plateau splitting at larger wavelengths.\;---}
As the source strength or the peak distance is increased further, the source term will increasingly deform the stationary profile compared to the pattern of the mass-conserving system.
The weak source terms create weak spatial gradients.
These gradients let long plateaus bend inward significantly toward $\rho_\mathrm{hss}$ [Fig.~\ref{fig:3}(b), see ~\cite{supplemental_material}].
When the maximum of the low-density plateau (or the minimum of a high-density plateau) enters the region of negative NC slope ${\partial_\rho\eta^*<0}$ at wavelengths ${\Lambda \geq \Lambda_\mathrm{split}}$, the plateau loses stability, and a new peak or mesa is nucleated [Fig.~\ref{fig:3}(b)] \cite{Kolokolnikov.etal2007,Kolokolnikov.etal2014,Brauns.etal2021}.
Thus, splitting and interrupted coarsening steer the pattern toward a stationary, periodic configuration with a wavelength $\Lambda$ fulfilling ${\Lambda_\mathrm{stop}<\Lambda<\Lambda_\mathrm{split}}$ [Fig.~\ref{fig:3}(c)].

\textit{Successive coalescence and splitting.\;---}
The scale separation ${\Lambda_\mathrm{stop}\ll \Lambda_\mathrm{split}}$ can be lost if the source term (approximately) vanishes at a plateau density $\rho_\pm$, as in the mKS model with a logistic growth term ${s(\rho)=\rho (p-\rho)}$.
Far below the splitting threshold $\Lambda_\mathrm{split}$, coalescence is counteracted by production ${\varepsilon s(\rho_-)\approx \varepsilon\rho_-}$ in the low-density plateau ${\rho_-\approx 0}$ [Eq.~\eqref{eq:interrupted-coarsening}; dotted line in Fig.~\ref{fig:3}(c)].
In contrast, near the splitting threshold, the plateau deformation inducing splitting is driven by a source term ${\sim\varepsilon \rho\gg \varepsilon \rho_-}$ because the plateau bends upward.
Effectively, the source term appears weak for interrupted coarsening but strong for splitting, causing both thresholds to converge already at small source strength $\varepsilon$.
Taking into account that the plateau density grows as the splitting threshold $\Lambda_\mathrm{split}$ is approached (see~\cite{supplemental_material}), the threshold of interrupted coarsening $\Lambda_\mathrm{stop}$ bends downward [solid blue line in Fig.~\ref{fig:3}(c)].
Nonetheless, the difference ${\Lambda_\mathrm{split}-\Lambda_\mathrm{stop}}$ decreases strongly with $\varepsilon$, well describing the numerical results in the mKS model [Fig.~\ref{fig:3}(c)].

As a result, at moderate source strength $\varepsilon$, several consecutive coalescence and splitting events occur before the stationary, periodic pattern develops [Fig.~\ref{fig:3}(d) top].
At larger source strength $\varepsilon$, the domain size would have to be fine-tuned for stationary periodic patterns to exist.
Instead, the pattern continuously undergoes coalescence and splitting, reaching a dynamic steady state that has been argued to be chaotic [Fig.~\ref{fig:3}(d) bottom] \cite{Painter.Hillen2011}.

\paragraph{Conclusions and outlook.\;---}
Taken together, we have studied the generic dynamics of chemotactic aggregates using the mass-redistribution potential $\eta$, whose gradients fully determine cell migration by integrating random and chemotactic movement.
Our geometric construction principle, akin to a Maxwell construction, leverages flux-balance and reactive area balance, directly linking to the underlying chemotactic physics.

The construction and macroscopic dynamics closely resemble those in two-component mass-conserving reaction--diffusion (2cMcRD) systems \cite{Otsuji.etal2007,Ishihara.etal2007,Halatek.Frey2018,Brauns.etal2020,Brauns.etal2021,Weyer.etal2023}, although pattern formation in KS models is driven by a nonlinear cross-diffusion term, producing a mass-redistribution potential that depends highly nonlinearly on the cell and chemoattractant densities.
We expect that the phenomena likewise extend to higher-dimensional systems \cite{Meyer.etal2014,OByrne.Tailleur2020,Weyer.etal2024}.
As equilibrium-like pattern constructions were also obtained for scalar active matter~\cite{Solon.etal2018,Solon.etal2018a,Tjhung.etal2018,Fausti.etal2021,Omar.etal2023,Langford.Omar2023,Cates.Nardini2023,Langford.Omar2024}, it is an exciting question how these constructions can be used to classify the coarse-grained dynamics of scalar non-equilibrium matter more generally.

We expect that our theory will offer valuable insights into multi-species chemotactic systems~\cite{Wolansky2002,Liu.etal2019a,Muramatsu.etal2025}, their interplay with other aggregation mechanisms such as phase separation~\cite{Zhao.etal2023,Demarchi.etal2023,Goychuk.etal2024}, and systems with more complex signaling~\cite{Ziepke.etal2022}.
In particular, interrupted coarsening and domain splitting may shape the thymus medulla \cite{Muramatsu.etal2025}.
The predicted phenomena may also be experimentally accessible using engineered bacteria (see, e.g., Ref.~\cite{Curatolo.etal2020}).
An intriguing open question is
whether chemorepelling systems \cite{Liebchen.etal2015} and traveling fronts in systems with chemoattractant consumption or nutrient uptake can be analyzed using a similar framework~\cite{Tindall.etal2008,Adler1966,Budrene.Berg1995,Cremer.etal2019}.
Finally, understanding how noise affects wavelength selection and spatiotemporally chaotic steady states remains an important challenge for future work.

\smallskip

\begin{acknowledgments}
We thank Fridtjof Brauns, Natan Dominko Kobilica, and Florian Raßhofer for inspiring discussions.
We thank Alexander Ziepke for his critical reading of the manuscript.
This work was funded by the Deutsche Forschungsgemeinschaft (DFG, German Research Foundation) through the Excellence Cluster ORIGINS under Germany’s Excellence Strategy – EXC-2094 – 390783311, the European Union (ERC, CellGeom, project number 101097810), and the Chan-Zuckerberg Initiative (CZI).
\end{acknowledgments}

\appendix

\section{Nullcline-slope criterion exact for short-wavelength perturbations}
\label{app:slope-criterion}
The nullcline-slope criterion predicts the stability of the hss against chemotactic aggregation.
It is derived above using a QSS approximation in the long-wavelength limit.
For short-wavelength perturbations resulting in fast mass redistribution, the QSS approximation does not include the timescale of the local reactive relaxation of $\eta$ onto $\eta^*(\rho)$.
Nevertheless, $\eta^*(\rho)$ still sets the relaxation direction, and the nullcline-slope criterion, Eq.\eqref{eq:slope-crit}, remains valid: it determines whether $\eta$ increases or decreases in regions of increased cell density $\rho$, and thus whether uniform densities are destabilized~\cite{Weyer.etalinpreparationa}.

\section{Quantifying coarsening of chemotactic aggregates}
\label{app:coarsening-quant}
To understand the dynamics of patterns with several domains (peaks or mesas), we describe the interaction between two elementary pattern motifs (two half-peak or two half-mesa patterns).
As illustrated for peaks in the following, the pattern domains can interact either via competition for mass or coalescence with neighbors (mirror image) [Fig.~\ref{fig:2}(a,b); reflective boundaries].
Assuming a slow evolution of the pattern after each \emph{single} elementary pattern relaxed to its (quasi-)stationary profile, we approximate the mass-redistribution potential at the single interfaces by its stationary value $\etaStat$ (QSS approximation).
Via the reactive area balance, $\etaStat$ depends on the extremal densities $\check{\rho}$ and $\hat{\rho}$ of the elementary pattern [Fig.~\ref{fig:2}(a,b)].
Consequently, differences in peak height or in the width of the upper and lower plateau regions generate $\eta$ gradients:
Changes in the plateau lengths shift $\check{\rho}$ or $\hat{\rho}$ exponentially, thereby altering $\eta_{\mathrm{stat}}$.
For peak patterns, where the upper plateau is absent, the peak mass directly sets the height $\hat{\rho}$ and thus determines $\eta_{\mathrm{stat}}$.
All such changes of the pattern reflect effective changes of the mass $M$ of the elementary pattern, since changes in plateau length due to a shift of the peak or mesa interface [cf.\ Fig.~\ref{fig:2}(b)] require mass changes ${\delta M \approx \pm (\hat{\rho} - \check{\rho}) \, \delta l}$; the sign distinguishes whether the upper or lower plateau is varied.
The corresponding partial derivatives of the stationary mass-redistribution potential are denoted by $\partial_M^{\pm} \eta_{\mathrm{stat}}$.
Since the geometry of the reactive area balance is model-independent, it generically implies ${\partial_{\hat{\rho},\check{\rho}}\etaStat \lessgtr 0}$ and, with the above definition of $\delta M$, ${\partial_M^{\pm} \eta_{\mathrm{stat}} < 0}$ for both peak and mesa patterns, assuming that $\hat{\rho}$ increases with mass in peaks [Fig.~\ref{fig:2}(a,b)].

In peak competition [Fig.~\ref{fig:2}(a)], the mass difference $\delta M$ of the two peaks generates an $\eta$ gradient ${\sim  \partial_M^+\etaStat/\Lambda}$ that drives mass redistribution; the larger peak grows, the smaller collapses.
In coalescence [Fig.~\ref{fig:2}(b)], unequal plateau widths shift $\etaStat$, inducing a gradient ${\sim  \partial_M^-\etaStat/\ell_\mathrm{int}}$ within the peak of half-width $\ell_\mathrm{int}$.
This gradient drives peak motion until the shorter plateau vanishes.
In both cases, the symmetric configuration is destabilized by a \emph{mass-competition instability}.
Its growth rates $\sigma_\pm$ are derived in the Supplemental Material~\cite{supplemental_material} within the QSS approximation and in Ref.~\cite{Weyer.etalinpreparationa} using singular perturbation theory.
Singular perturbation theory shows that the instability (coarsening) criteria $\partial_M^\pm\eta_\mathrm{stat}<0$ remain exact beyond the QSS approximation.

In terms of diffusion and chemotaxis, what causes generic aggregate coarsening analogously to Ostwald ripening?
Linear stability of the low-density plateau requires ${\partial_{\rho} \eta^*|_{\rho_-} > 0}$ and ${\partial_c f(\rho_-, c_-) < 0}$, which together imply ${\partial_\rho c^*|_{\rho_-} = -\partial_\rho f / \partial_c f|_{\rho_-} > 0}$.
Accordingly, ${\partial_M^+ \eta_{\mathrm{stat}} < 0}$ implies ${\partial_M^+ \rho_- < 0}$, such that cells diffuse toward larger aggregates because these deplete the surrounding low-density plateau more strongly.
In contrast, chemotaxis drives cells toward smaller aggregates since ${\partial_M^+ c_- < 0}$, but the generic finding ${\partial_M^{\pm} \eta_{\mathrm{stat}} < 0}$ implies that the diffusive flux dominates.
Taken together, although chemotaxis biases motion of cells towards smaller aggregates, their random motion dominates in the surrounding low-density plateaus and leads to the growth of larger aggregates.
These deplete their surroundings more strongly, resulting in aggregate coarsening.

\section{Coarsening in the mKS model}
\label{app:coarsening-mKS}
In the mKS model, we find  ${\sigma^\pm \sim \exp(-\Lambda/\gamma_\pm)}$ where $\Lambda$ is the wavelength of the pattern [Fig.~\ref{fig:2}(a,b)] and $\gamma_\pm$ are length scales related to the exponential approach of the pattern profile towards $\rho_\pm$ (see~\cite{supplemental_material} and Ref.~\cite{Weyer.etalinpreparationa}).
One has ${\gamma_-\gtrless \gamma_+}$ for ${\Bar{\rho} \gtrless 2 D_\rho/T}$ [Fig.~\ref{fig:2}(d), inset].
Consequently, coalescence is faster for large average densities ${\Bar{\rho} > 2 D_\rho/T}$ (competition at low densities ${\Bar{\rho} < 2 D_\rho/T}$), and dominates the coarsening process at large wavelengths.
This prediction agrees with the observations in numerical kymographs [Fig.~\ref{fig:2}(c)].
As ${[\sigma^\pm(\langle\Lambda\rangle)]^{-1}}$ is the typical timescale on which peaks vanish via competition or coalescence at the mean pattern wavelength $\langle\Lambda\rangle$ (and strongly increases with $\langle\Lambda\rangle$), the time during coarsening scales as ${t^{-1} \sim \sigma^\pm(\langle\Lambda\rangle)}$ \cite{Langer1971,Glasner.Witelski2003,Weyer.etal2023}; see~\cite{supplemental_material}. 
The resulting coarsening law ${\langle\Lambda\rangle\sim \max(\gamma_\pm) \log t}$ describes numerical simulations well, with deviations at the transition from competition- to coalescence-dominated coarsening [Fig.~\ref{fig:2}(d)].

\section{Mass competition under the influence of weak source terms}
\label{app:IC}
The effect of the source term onto the coalescence of neighboring peaks [Fig.~\ref{fig:3}(a)] can be analyzed by making a traveling-wave ansatz ${\rho(x,t) = \tilde{\rho}(x-\delta x(t))}$ for the aggregate.
Integrating Eq.~\eqref{eq:cont-eq} from the peak maximum to the right boundary, one finds ${(\hat{\rho}-\rho_-)\partial_t^\mathrm{source}\delta x \approx -\varepsilon s(\rho_-)\delta x}$, which drives the peak back to its symmetric position ${\delta x = 0}$ (see~\cite{supplemental_material}).
Moreover, to lowest order in $\varepsilon$, the destabilizing mass-competition effect due to $\eta$ gradients within the peak is described by the result $\partial_t^\mathrm{comp}\delta x \approx \sigma^-_\mathrm{D} \delta x$ for the mass-conserving system.
Together, one obtains the growth rate of the mass-competition instability under the influence of weak source terms (for ${\varepsilon \ll 1}$; cf.\ Refs.~\cite{Kang.etal2007,Kolokolnikov.etal2014,Kong.etal2024} and \cite{Kolokolnikov.etal2006,Brauns.etal2021,Weyer.etal2023} for the analogous effect in two-component reaction--diffusion systems)
\begin{equation}\label{eq:interrupted-coarsening}
    \sigma_\varepsilon^-(\Lambda) \approx \sigma_\mathrm{D}^-(\Lambda) - \varepsilon \frac{s(\rho_-)}{\hat{\rho}(\Lambda)-\rho_-}.
\end{equation}
Here, $\sigma_\mathrm{D}^-(\Lambda)$ denotes the diffusion-limited part of the growth rate $\sigma^-$.
A detailed derivation is given in the Supplemental Material~\cite{supplemental_material}.

\end{document}


\maketitle

\tableofcontents

\clearpage

\section{The mass-redistribution potential}
In the main text, we define the mass redistribution potential
\begin{equation} 
\label{eq:SI_eta_def}
\eta \equiv \frac{D_\rho}{T}\eta_\rho - \eta_c\, 
\end{equation}
using
\begin{equation} 
\eta_\rho(\rho) \equiv \int_{\rho_0}^{\rho} \mathrm{d}\rho'\, \frac{1}{\chi_\rho(\rho') \rho'}\, , \qquad  \eta_c(c)\equiv \int_{c_0}^{c} \mathrm{d}c'\, \chi_c(c')
\, , 
\end{equation}
where we chose arbitrary reference densities $\rho_0$ and $c_0$.
Under the assumptions ${\chi_\rho,\chi_c>0}$ and ${\rho,c > 0}$, the functions $\eta_\rho(\rho)$ and $\eta_c(c)$ are invertible.
Thus, either two of the three variables $\rho$, $c$, and $\eta$ can be chosen to be independent.
The third one then follows as a function of the others:
\begin{subequations}\label{eq:density-expressions}
    \begin{align}
        \eta(\rho,c) &= \frac{D_\rho}{T} \eta_\rho - \eta_c,\\
        c(\rho, \eta) &= \eta_c^{-1}\left(\frac{D_\rho}{T} \eta_\rho-\eta\right),\label{eq:c-expression}\\
        \rho(c, \eta) &= \eta_\rho^{-1}\left(\frac{T}{D_\rho} (\eta+\eta_c)\right).\label{eq:rho-expression}
    \end{align}
\end{subequations}

The mass-redistribution potential $\eta$ follows its own dynamic equation.
It can be derived from the definition of the mass-redistribution potential Eq.~\eqref{eq:SI_eta_def} together with the generalized Keller--Segel equations [Eq.~(2) and the chemoattractant equation in the main text] and reads
\begin{equation}
    \partial_t \eta 
    = 
    \frac{D_\rho}{T} \frac{1}{\chi_\rho \rho} \, 
    \partial_t \rho 
    - \chi_c \, \partial_t c
    \, .
\end{equation}
Thus, one obtains in the mass-conserving case ${\varepsilon=0}$
\begin{align} 
\label{eq:SI_eta_eom}
\partial_t \eta &= \frac{D_\rho}{T} \frac{\partial_t \rho}{\chi_\rho \rho} + D_c\chi_c\nabla \frac{1}{\chi_c}\nabla \eta - \chi_c\frac{D_\rho D_c}{T}\nabla \frac{1}{\chi_c} \nabla\eta_\rho - \chi_c f(\rho_, c)\, , \\
\label{eq:KS-eta}
&= \frac{D_\rho}{\chi_\rho \rho} \nabla \chi_\rho \rho \nabla \eta + D_c \chi_c \nabla \frac{1}{\chi_c} \nabla \eta - \chi_c \frac{D_\rho D_c}{T} \nabla \frac{1}{\chi_c \chi_\rho \rho} \nabla \rho - \chi_c \Tilde{\Tilde{f}}(\rho, \eta)\, . 
\end{align}
Here, we defined ${\Tilde{\Tilde{f}}(\rho, \eta)\equiv f(\rho,c(\rho,\eta))}$ using Eq.~\eqref{eq:c-expression}.
These dynamics contain two (generalized) diffusion terms which vanish for constant $\eta$.
Moreover, the last two terms act as source terms that increase or decrease the mass-redistribution potential depending on the cell-density profile $\rho$.

\clearpage\newpage

\section{Reactive equilibria}
\label{sec:local-dynamics}

In a well-mixed system, all gradients vanish and the Keller--Segel dynamics reduces to
\begin{subequations}\label{eq:local-dyn}
    \begin{align}
        \partial_t \rho 
        &= \varepsilon s(\rho) 
        \, ,\\
        \partial_t c 
        &= f(\rho,c)
        \, .
    \end{align}
\end{subequations}
We call this the \emph{local dynamics} because it occurs locally at each point of the domain while the other terms of the Keller--Segel dynamics describe the diffusive and chemotactic redistribution of the densities, which couple different regions of the domain.

\paragraph{Mass-conserving dynamics.\;---}
In the mass-conserving systems (${\varepsilon=0}$), the cell density remains constant ${\rho = \Bar{\rho}}$ as set by the initial condition, and only the chemoattractant concentration equilibrates to the reactive equilibrium $c^*$ defined by
\begin{equation}
    f(\bar\rho,c^*)=0.
\end{equation}
Equivalently, the mass-redistribution potential evolves by $\partial_t \eta = - \chi_c \Tilde{\Tilde{f}}(\bar\rho,\eta)$ towards the reactive equilibrium $\eta^*$.

The family of reactive equilibria $f=0$ for different densities $\rho$ is called the nullcline (NC).
If the reactions exhibit multistability, several reactive equilibria exist for a fixed density $\rho$.
We denote the nullcline by $c^*(\rho)$ [and $\eta^*(\rho)$, $\eta^*(c)$].
In the case of multistable local reactions, several branches of the nullclines $c^*(\rho),\eta^*(\rho)$ have to be defined.
Importantly, with Eq.~\eqref{eq:rho-expression}, the condition ${\partial_\rho f>0}$ implies for ${\Tilde{f}(c, \eta)\equiv f(\rho(c,\eta),c)}$ that one has ${\partial_\eta \tilde{f}(c,\eta)>0}$.
Thus, the nullcline $\eta^*(c)$ in $(c,\eta)$ phase space is necessarily  a single-valued function.

\paragraph{Broken mass conservation.\;---}
If mass conservation is broken by a finite growth and death rate $\varepsilon>0$, a (discrete set of) steady state(s) is selected by the intersection(s) of the two nullclines $f=0$ and $s=0$.
The density of the chemotactic species $\rho$ is set by the balance of production and degradation $s=0$ instead of the initial condition.

\clearpage\newpage

\section{Stationary patterns of the mass-conserving Keller--Segel dynamics}
\label{sec:stat-patterns}

In this section, we give derive the reactive area balance resulting in the generalized Maxwell construction for stationary patterns of the mass-conserving Keller--Segel dynamics ($\varepsilon=0$).
This construction of the one-dimensional pattern profiles can be used to determine the dynamics of patterns in higher-dimensional systems assuming that the pattern interfaces are narrow compared to the curvature radii of the interfaces \cite{Bray2002,Pismen2006,Brauns.etal2021,Weyer.etalsubmitted}.

\subsection{Reactive area balance}
\label{sec:reactive-area-balance}
In the main text, we argue that the value of the stationary mass-redistribution potential $\etaStat$ is fixed by the constraint that the overall production and degradation of the chemoattractant balance.
To derive this balance, we consider the stationary chemoattractant equation (\emph{chemoattractant-profile equation})
\begin{equation}\label{eq:profile-eq-c}
    0 
    = 
    D_c \partial_x^2 c_\mathrm{stat} 
    + 
    \Tilde{f}(c_\mathrm{stat},\etaStat)
    \, .
\end{equation}
Here, we introduced ${\Tilde{f}(c, \eta)\equiv f(\rho(c,\eta),c)}$ using Eq.~\eqref{eq:rho-expression}.
No-flux or periodic boundary conditions for the chemoattractant imply that integrating Eq.~\eqref{eq:profile-eq-c} over the whole domain yields
\begin{equation}
    0 = \int_0^\frac{\Lambda}{2}\mathrm{d}x\, \Tilde{f}(\cStat,\etaStat)\, .
\end{equation}
This condition describes the balance of production and degradation, which are both contained in the reaction term $f$.
The balance condition can be evaluated in phase space by first multiplying Eq.~\eqref{eq:profile-eq-c} by $\partial_x \cStat$.
Thereby, one derives the \emph{total turnover balance} condition (cf.~Ref.~\cite{Brauns.etal2020} for the analogous condition in two-component mass-conserving reaction--diffusion systems)
\begin{equation}\label{eq:ttb}
    0 = \int_{\check{c}}^{\hat{c}}\mathrm{d}c\, \Tilde{f}(c,\etaStat) 
    = 
    \int_{\check{\rho}}^{\hat\rho}
    \mathrm{d}\rho \, 
    \frac{\Tilde{\Tilde{f}}(\rho, \etaStat)}{\chi_c\chi_\rho \rho}
    \, .
\end{equation}
Here, the second expression in terms of $\rho$ follows as a reparametrization using Eq.~\eqref{eq:c-expression}.
The densities $\check{\rho},\check{c}$ and $\hat{\rho},\hat{c}$ are the cell and chemoattractant densities of the elementary stationary pattern at the left and right domain boundary.
Demanding monotonicity of the elementary patterns, those are defined as the minimal and maximal densities.\footnote{%
Whenever the stationary pattern contains an internal minimum or maximum, both $\partial_x\rhoStat$ and $\partial_x\cStat$ vanish there [cf.~Eq.~\eqref{eq:rho-expression}], and one can split the pattern into monotonous elementary patterns on separate domains with no-flux boundary conditions.
If a periodic pattern is constructed from concatenating and reflecting different elementary patterns, these are typically unstable, for example, see Ref.~\cite{Ward.Wei2002} for an example in reaction--diffusion systems.
We do not consider those patterns here.
}
The total turnover balance condition Eq.~\eqref{eq:ttb} represents a (generalized) area balance in $(c,\eta)$ phase space.
To show this, we observe that Eq.~\eqref{eq:rho-expression} implies
\begin{equation}
    \partial_\eta \tilde{f} = (\partial_\rho f) \frac{T}{D_\rho}\rho \chi_\rho.
\end{equation}
With this, we rewrite the total turnover balance Eq.~\eqref{eq:ttb} as the reactive area balance
\begin{equation}\label{eq:reactive-area-balance-c}
    0 = \int_{\check{c}}^{\hat{c}}\mathrm{d}c\,\int_{\eta^*(c)}^{\etaStat}
    \mathrm{d}\eta \, 
    \frac{T}{D_\rho} \rho \chi_\rho
    \partial_\rho f
    \, .
\end{equation}
The double integral covers the areas between the NC and FBS in $(c,\eta)$ phase space from the minimal to the maximal chemoattractant density of the pattern [cf.\ Fig.~1(c,d)].
The integrand ${T\rho \chi_\rho \partial_\rho f/D_\rho > 0}$ is positive by assumption and can thus be interpreted as the surface element of a nontrivial metric with respect to which the surface area between NC and FBS is measured.
To derive the reactive area balance in $(\rho,\eta)$ phase space, we note that Eq.~\eqref{eq:c-expression} gives
\begin{equation}
    \partial_\eta \tilde{\tilde f} = - \frac{\partial_c f}{\chi_c}\, .
\end{equation}
The total turnover balance Eq.~\eqref{eq:ttb} can then be rewritten as
\begin{equation}\label{eq:reactive-area-balance-rho}
    0 = \int_{\check{\rho}}^{\hat{\rho}}\mathrm{d}\rho\,\int_{\eta^*(\rho)}^{\etaStat}\mathrm{d}\eta\, \frac{1}{\chi_c(c(\rho,\etaStat))\chi_\rho \rho} \frac{-\partial_c f(\rho,c(\rho,\eta))}{\chi_c(c(\rho,\eta))} .
\end{equation}
Note that the two $\chi_\mathrm{c}$ in the denominators are evaluated at two different values of $c$.
Here, the integrand is again positive and may be interpreted as a surface element if one has $\partial_c f<0$.
This is necessarily fulfilled close to the nullcline if the local reactive dynamics is stable (see Sec.~\ref{sec:local-dynamics}).
Then, the total turnover balance is equivalent to a reactive area balance in $(\rho,\eta)$ phase space, 
again with a nontrivial metric.

\paragraph{Construction in $(c,\eta)$ phase space.\;---}
If the derivative $\partial_c f$ changes sign at a distance from the nullcline or in the case of multistability of the local chemoattractant dynamics (cf.\ Sec.~\ref{sec:local-dynamics}), the turnover balance only strictly corresponds to an area balance in the $(c,\eta)$ phase space.
In $(\rho,\eta)$ phase space, the integrand of the integral Eq.~\eqref{eq:reactive-area-balance-rho}, that is, the surface element becomes negative in some regions of phase space.
The construction can then be performed in $(c,\eta)$ phase space, and the cell density is reconstructed using Eq.~\eqref{eq:rho-expression}.
In particular, Eq.~\eqref{eq:rho-expression} implies that the cell density $\rhoStat$ varies monotonously with $\cStat$.

In the following, we show that the classification into peak and mesa patterns based on the nullcline shape in $(c,\eta)$ phase space is equivalent to the classification $(\rho,\eta)$ phase space.
To this end, we will show that the two nullclines ${\eta^*(\rho)}$ and ${\eta^*(c)}$ are both either $\mathsf{\Lambda}$- or $\mathsf{N}$-shaped.
First, also in the case of multistability of the reaction kinetics, the nullcline $\eta^*(c)$ in $(c,\eta)$ phase space is single-valued (see Sec.~\ref{sec:local-dynamics}).
Second, note that, using Eqs.~\eqref{eq:density-expressions}, one has
\begin{equation}\label{eq:rho-c-slope-rel}
    (\partial_\rho f) \partial_c \eta^* = - (\partial_c f) \partial_\rho \eta^*
\end{equation}
and by assumption ${\partial_\rho f>0}$.
This implies that any extremum of ${\eta^*(\rho)}$ is an extremum of ${\eta^*(c)}$.
In contrast, assuming that one has ${\partial_c \eta^*=0}$ but ${\partial_\rho \eta^*\neq 0}$, it follows that ${\partial_c f=0}$.
However, it also holds that [see Eqs.~\eqref{eq:density-expressions}]
\begin{equation}\label{eq:rho-slope-expr}
    \frac{\partial_c f}{\chi_c}\partial_\rho\eta^* = \partial_\rho f + \frac{D_\rho}{T\rho\chi_\rho\chi_c}\partial_c f.
\end{equation}
Thus, ${\partial_c f=0}$ implies ${\partial_\rho\eta^*=\infty}$ and ${\partial_c \eta^*=-\chi_c}$, in contrast to the assumption ${\partial_c \eta^*=0}$.
Taken together, ${\eta^*(\rho)}$ has an extremum if and only if ${\eta^*(c)}$ has an extremum, and both nullclines are either $\mathsf{\Lambda}$- or $\mathsf{N}$-shaped.
Consequently, the classification of the stationary patterns into peak and mesa patterns can be performed in either of the two [$(c,\eta)$ and $(\rho,\eta)$] descriptions.

In summary, the flux-balance condition ${\etaStat =\mathrm{const.}}$ together with the reactive area balance Eqs.~\eqref{eq:reactive-area-balance-rho},~\eqref{eq:reactive-area-balance-c} correspond to a generalized Maxwell construction in the $(\rho,\eta)$ and $(c,\eta)$ phase spaces.

\begin{figure}[h]
    \centering
    \includegraphics[width=\textwidth]{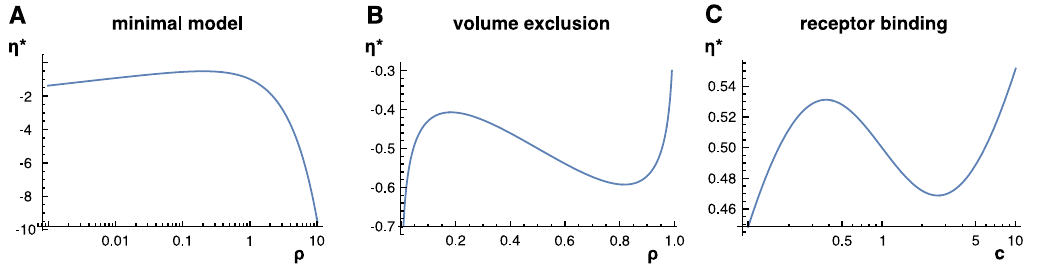}
    \caption{Nullclines for the minimal Keller--Segel model (a), in the model with volume exclusion (b), and considering receptor-binding kinetics (c).
    The parameter are chosen as $\rho_\mathrm{max}=1$, $D_\rho/T=0.2$ in (a) and (c), and $D_\rho/T=0.15$ in (b).}
    \label{fig:KS-nullclines}
\end{figure}

\paragraph{Examples of pattern classification.\;---}
In the main text we show that, with this generalized Maxwell construction, the elementary stationary patterns can be classified as peak and mesa patterns, determined by the shape of the nullcline $\eta^*(\rho)$.
Figure~\ref{fig:KS-nullclines} shows the nullclines for several classical examples of Keller-Segel models.
The minimal Keller--Segel (mKS) model \cite{Childress.Percus1981} has the $\mathsf{\Lambda}$-shaped nullcline $\eta^* (\rho) = \frac{D_\rho}{T} \log(\rho) - \rho$ and shows peak patterns (Fig.~\ref{fig:KS-nullclines}a).
The model with volume exclusion \cite{Hillen.Painter2001, Painter.Hillen2002}, which introduces a maximal cell density $\rho_\mathrm{max}$, has the \textsf{N}-shaped nullcline ${\eta^*(\rho) = \frac{D_\rho}{T}\log[\rho/(\rho_\mathrm{max}-\rho)]-\rho+\mathrm{const.}}$ (Fig.~\ref{fig:KS-nullclines}b).
Finally, the sensitivity $\chi_c$ becomes nontrivial when considering receptor-binding kinetics \cite{Segel1977,Stevens.Othmer1997}.
A simple form accounting for receptor saturation gives ${\chi_c = K/(K+c)^2}$.
This leads to a \textsf{N}-shaped nullcline as well that, parameterized by the chemoattractant density $c$ and for linear reactions ${f=\rho-c}$, has the form ${\eta^*(c) = \frac{D_\rho}{T} \log(c) - K/(K+c)}$ (Fig.~\ref{fig:KS-nullclines}c).
We give details of the construction for both pattern types below.

\subsection{Plateau stability}
\label{sec:plateau-stability}

In the main text, we argued that plateau regions and inflection points of the stationary pattern correspond to FBS-NC intersections.
The question remains which conditions determine whether intersections give inflection points or describe plateau regions.
The key observation to distinguish both cases is that only those FBS-NC intersections correspond to plateaus of the pattern that are both locally and laterally stable.
Otherwise, an approximately flat pattern region at those densities would be unstable against small uniform or spatially varying perturbations, breaking up the plateau.
In systems with locally stable reaction kinetics (cf.\ Sec.~\ref{sec:local-dynamics}), that is, without multistability, local stability is ensured by assumption and lateral stability implies that the nullcline slope at these intersections must be positive ${\partial_\rho\eta^*>0}$ [cf.\ Eq.~(4) in the main text].
If the reaction kinetics shows multistability, one must additionally ensure local stability ${\partial_c f < 0}$.
Both conditions ${\partial_c f < 0}$ and ${\partial_\rho\eta^*(\rho)>0}$ together imply ${\partial_c\eta^*(c)>0}$.
Conversely, ${\partial_c\eta^*(c)>0}$ implies that either these two conditions hold or ${\partial_c f > 0}$ and ${\partial_\rho\eta^*(\rho)<0}$ [see Eq.~\eqref{eq:rho-c-slope-rel}].
However, the second case must not occur because it contradicts the relation Eq.~\eqref{eq:rho-slope-expr}.
Consequently, ${\partial_c\eta^*(c)>0}$ is a necessary and sufficient criterion for local and lateral stability.
As a result, in the case of multistability, the geometric construction of the stationary pattern can be performed in $(c,\eta)$ phase space.
FBS-NC intersections with a positive nullcline slope ${\partial_c\eta^*(c)>0}$ correspond to plateau regions of the pattern while an intersection with negative nullcline slope ${\partial_c\eta^*(c)<0}$ corresponds to an inflection point of the pattern.
For locally stable reaction kinetics, the same classification applies to the nullcline $\eta^*(\rho)$ in $(\rho,\eta)$ phase space.

\clearpage\newpage

\section{Mass-competition instability in the mass-conserving Keller--Segel dynamics}
\label{sec:mass-comp}

In this section, we derive the rates of the mass-competition instability for peak coalescence and peak competition using a QSS approximation for the mass-redistribution potential at each single peak (interface).
We derive the full growth rates including the timescale of local reactive relaxation using a singular perturbation analysis in Ref.~\cite{Weyer.etalinpreparationa}, where we also give the analogous expressions for competition and coalescence of mesa patterns.
Importantly, in both the diffusion- and reaction-limited regimes, the instability criterion is the same.
In both cases, the mass-competition instability is driven by a decrease in the stationary mass-redistribution potential with the (associated) domain mass (${\partial_M^\pm \etaStat < 0}$).
After the derivation of the general expressions, we apply these rates to the mKS model and derive coarsening exponents from them.

\subsection{The mass-competition rates in the quasi-steady-state limit}
\label{sec:mass-comp-QSS}

Periodic patterns of peaks can be constructed from concatenating and reflecting elementary patterns and, thus, are stationary patterns.
Importantly, in the main text, we discuss that peaks compete for mass and undergo coalescence, driving a coarsening process of patterns with several peaks.
To analyze these two coarsening modes (peak competition and coalescence) we segment the periodic pattern into symmetric patterns 
comprising one wavelength of the periodic pattern (constructed from two elementary stationary patterns) on a domain $[-\Lambda/2,\Lambda/2]$ of length $\Lambda$ with reflective boundary conditions \cite{Brauns.etal2021,Weyer.etal2023}.
The resulting pattern consists of either a peak centered at the middle of the domain [Fig.~2(b) in the main text] whose coalescence with one of the reflective boundaries corresponds to the coalescence with the left- or right-neighboring peak in the periodic pattern, or two half-peaks at either boundary of the domain that compete for mass [Fig.~2(a) in the main text].
Because mass redistribution is faster on shorter distances, we only consider the competition of neighboring peaks of the periodic pattern.

For a small mass difference $\delta M$ [cf.\ Fig.~2(a) in the main text] between the half peaks or a small deviation of the peak position from the domain middle $\delta x$ [cf.\ Fig.~2(b) in the main text], competition and coalescence dynamics are quantified by the growth rates $\sigma_\mathrm{D}^\pm$ of these deviations \cite{Brauns.etal2021,Weyer.etal2023}.
We first consider the peak-competition scenario [Fig.~2(a) in the main text].
Assuming that changes in the plateau densities are small compared to the mass $\delta M$ transferred from one peak to the other, the profile in the pattern plateaus between the peaks relaxes quickly.
Thus, the pattern plateau can be approximated by a stationary state.
In the stationary state, the continuity equation, Eq.~(2) with ${\varepsilon=0}$ in the main text, and the approximately uniform density $\rho\approx\rho_-$ implying ${\chi_\rho\rho \approx \chi_\rho(\rho_-)\rho_-}$, yield ${\partial_x^2\eta\approx 0}$, i.e., a linear spatial profile for the mass-redistribution potential $\eta(x)$.
Moreover, given that mass competition for well-separated peaks is slow compared to the relaxation of the mass-redistribution potential at each individual peak, we approximate $\eta$ at the peaks by its stationary value ${\etaStat(\bar{M}\pm\delta M)\approx \etaStat(\bar{M})\pm\delta M \partial_M^+\etaStat}$ (QSS approximation).
Here, the peak masses are ${M_{1,2} = \bar{M}\pm\delta M}$ [cf.\ discussion in the main text].
Thus, during mass competition, the linear slope of $\eta$ is $2\partial_M^+\etaStat/\Lambda$, 
neglecting the peak width $\ell_\mathrm{int}$ compared to the peak separation $\Lambda$ (sharp-interface approximation).
Integrating the continuity equation [cf.\ Eq.~(2) for ${\varepsilon=0}$ in the main text] over the half-domain $[0,\Lambda/2]$, the mass change of the half-peak at ${x = \Lambda/2}$ is due to the mass fluxes at the two boundaries ${x = 0}$ and ${x = \Lambda/2}$.
Due to the reflective domain boundary, the mass flux is zero at ${x =\Lambda/2}$.
At ${x = 0}$ in contrast  the linear gradient in $\eta(x)$ leads to a mass flux that redistributes mass between the peaks changing the peak masses and thus the mass difference $\delta M$ as
\begin{equation}\label{eq:sigma-comp-QSS}
    \partial_t \delta M 
    \approx 
    - \frac{2 T \chi_\rho(\rho_-)\rho_-}{\Lambda/2} \, 
    (\partial_M^+ \eta_\mathrm{stat}|_{\Bar{M}}) \, 
    \delta M 
    \equiv 
    \sigma_\mathrm{D}^+ \, \delta M
    \, .
\end{equation}
As this is a linear equation for the growth of the deviation $\delta M$ with time, the growth rate is read off as the prefactor on the right hand side.
The subscript D denotes the diffusion-limited regime in which the competition rate is set by the mass transport between the peaks while we assume the relaxation of each single peak onto its QSS is fast in comparison (see Ref.~\cite{Weyer.etalinpreparationa}).
While the fraction $2\partial_M^+\etaStat/\Lambda$ describes the strengths of the induced gradient $\eta(x)$, the mobility in the low-density plateau $T\chi_\rho(\rho_-)\rho_-$ relates the gradient to the induced mass flux [Eq.~(2) in the main text].

Similarly, during coalescence [Fig.~2(b) in the main text], the rate of change in the peak displacement $\delta x$ [corresponding to a mass change ${\delta M = (\hat{\rho}-\check{\rho})\delta x}$] reads
\begin{equation}\label{eq:sigma-coal-QSS}
    \partial_t \delta x 
    \approx 
    - \frac{2 T \chi_\rho(\hat{\rho})(\hat{\rho}-\rho_-)}{\ell_\mathrm{int}} \, 
    (\partial_M^-\eta_\mathrm{stat}|_{\Bar{\rho}}) \, 
    \delta x
    \equiv 
    \sigma_\mathrm{D}^- \, \delta x
    \, .
\end{equation}
Thus, the coalescence rate $\sigma_\mathrm{D}^-$ is set by the strength of the induced gradient $\eta(x)$ given by $\partial_M^-\etaStat/\ell_\mathrm{int}$ and the mobility $T\chi_\rho(\rho_+)\rho_+$ evaluated at the (maximum) peak density because mass is transported through the peak from one peak interface to the other.
The mathematical derivation using singular perturbation theory in Ref.~\cite{Weyer.etalinpreparationa} accounts for the nonlinear $\eta$-profile in the peak and gives an explicit expression for the half-peak width $\ell_\mathrm{int}$.
Moreover, it yields the effect of a finite reactive relaxation at the single peaks, the effect of which we shortly summarize below.

The chemoattractant's finite production and degradation rate (set by the rates in the reaction term $f$) induces a timescale on which the mass-redistribution potential relaxes onto its QSS value $\eta_\mathrm{stat}$ at the peaks.
The mathematical analysis in Ref.~\cite{Weyer.etalinpreparationa} employing singular perturbation theory shows that this relaxation is described by the reaction-limited rates $\sigma_\mathrm{R}^\pm$ that limit the growth rates of the deviations $\delta M$, $\delta x$ if mass transport between the elementary patterns is instantaneous.
Because the reactive relaxation is driven by the changed stationary mass-redistribution potentials as well, one finds again ${\sigma_\mathrm{R}^\pm\sim -\partial_M^\pm\eta_\mathrm{stat}}$. 
The analysis in Ref.~\cite{Weyer.etalinpreparationa} also shows that the growth rates $\sigma^\pm$ of the mass-competition process incorporate both diffusive redistribution and reactive relaxation and follow from adding both timescales, i.e., ${(\sigma^\pm)^{-1}= (\sigma^\pm_\mathrm{D})^{-1} + (\sigma^\pm_\mathrm{R})^{-1}}$ (see Ref.~\cite{Weyer.etal2023} for the analogous effect in two-component mass-conserving reaction--diffusion systems).
Importantly, one has ${\sigma^\pm\sim -\partial_M^\pm\eta_\mathrm{stat}}$ such that the criterion for the mass-competition instability ${\partial_M^\pm\eta_\mathrm{stat} < 0}$ remains the same as derived under the QSS approximation here.

\subsection{Mass competition and coarsening in the minimal Keller-Segel Model}
\label{sec:mass-comp-mKS}
As an example, we explicitly calculate the peak-competition rate for the mKS model using the stationary peak profiles derived in Ref.~\cite{Kang.etal2007}.
Along with the peak-coalescence rate from the same reference [Eq.(2.20) for  ${\Lambda \gg 1}$], we will use these rates to determine the coarsening law that describes the evolution of the average pattern wavelength $\langle \Lambda \rangle$ in a large system containing many peaks.
The coarsening law is obtained using the scaling argument put forward in Refs.~\cite{Glasner.Witelski2003,Langer1971,Weyer.etal2023}.

In Ref.~\cite{Kang.etal2007}, the stationary peak profiles of the mKS model are obtained using asymptotic matching.
The cell-density profile $\rhoStat$ of a peak centered at ${x = 0}$ in the inner peak region [${x = \mathcal{O}(1/M)}$] is
\begin{equation}\label{eq:mKS-stat-rho}
    \rhoStat^\infty(x) \approx \frac{M^2 T}{8 D_\rho D_c} \operatorname{sech}^2\left(\frac{M T}{4 D_\rho D_c}x\right).
\end{equation}
Moreover, the corresponding chemoattractant density follows
\begin{equation}\label{eq:mKS-stat-c-inner}
    \cStat^\infty(x) \approx - \frac{D_\rho}{T} \log\left[4 \cosh^2\left(\frac{M T}{4 D_\rho D_c}x\right)\right] + \frac{M}{2 \sqrt{D_c}}\, .
\end{equation}
In the outer region [$x\gg \mathcal{O}(1/M)$], the chemoattractant density is given by
\begin{equation}\label{eq:mKS-stat-c-outer}
    \cStat^\infty(x) 
    \approx 
    \frac{M}{2 \sqrt{D_c}} \,
    \exp\left(-x/\sqrt{D_c}\right)
    \, .
\end{equation}
Determining the stationary mass-redistribution potential $\etaStat^\infty$ in the inner peak region using its definition $\eta = (D_\rho/T)\log\rho - c$ [cf.\ Eq.~(3) in the main text] and Eqs.~\eqref{eq:mKS-stat-rho},~\eqref{eq:mKS-stat-c-inner}, one finds
\begin{equation}\label{eq:mKS-stat-eta}
    \etaStat^\infty 
    \approx 
    \frac{D_\rho}{T} \log \left(\frac{M^2 T}{2 D_\rho D_c}\right) - \frac{M}{2 \sqrt{D_c}}
    \, .
\end{equation}

\subsubsection{Peak competition}

For the mKS model, Eq.~\eqref{eq:sigma-comp-QSS} yields the (diffusion-limited) peak-competition rate
\begin{equation} 
\label{eq:si_coars_rate_peak_comp}
        \sigma^+_\mathrm{D} 
        \approx  
        -\frac{4 T \operatorname{e}^{\frac{T}{D_\rho} \etaStat} \partial_{M}^{} \etaStat}{\Lambda} 
        = 
        - \frac{4 D_\rho}{\Lambda} \partial_M e^{\frac{T}{D_\rho} \etaStat}\, ,
\end{equation}
because it holds ${\etaStat = (D_\rho/T) \log \rho_- - \rho_-\approx (D_\rho/T) \log \rho_-}$ as $\rho_-$ is a FBS-NC intersection [cf.\ Fig.~1(c) in the main text].
Using the value of the stationary mass-redistribution potential Eq.~\eqref{eq:mKS-stat-eta}, one obtains 
\begin{equation} 
    \operatorname{e}^{\frac{T}{D_\rho}\eta_\mathrm{stat}} 
    \approx 
    \frac{T M^2}{2D_\rho D_c} 
    \operatorname{e}^{-\frac{TM}{2 D_\rho \sqrt{D_c}}} 
    \, .
\end{equation}
Inserting this into Eq.~\eqref{eq:si_coars_rate_peak_comp} then yields
\begin{equation} 
    \sigma_\mathrm{D} 
    \approx 
    \frac{4 T M }{\Lambda D_c} \left(\frac{T M }{4D_\rho \sqrt{D_c}} - 1\right)
    \exp\left(-\frac{TM}{2 D_\rho \sqrt{D_c}}\right) 
    \, .
\end{equation}
In a periodic pattern, the peak mass is determined by the average density and the peak's domain length, $M \approx \Lambda\bar\rho$ because in the minimal Keller--Segel model one has ${\rho_-\approx 0}$.
With this, one finds
\begin{equation}\label{eq:mKS-sigma-comp}
    \sigma^+_\mathrm{D} 
    \approx 
    \frac{4T\bar\rho}{D_c}
    \left(
    \frac{T\bar\rho \Lambda}{4 D_\rho\sqrt{D_c}} - 1
    \right) \, 
    \exp
    \left(
    -\frac{T\bar\rho}{2 D_\rho \sqrt{D_c}} \Lambda
    \right) 
     \, .
\end{equation}

\subsubsection{Coarsening in the minimal Keller--Segel model}

The growth rates for peak competition and coalescence allow us to make use of a scaling argument \cite{Weyer.etal2023,Langer1971,Glasner.Witelski2003} in order to obtain the dynamics of the average length scale $\langle\Lambda\rangle(t)$ of the pattern on a large domain.
Because on a large domain, both competition and coalescence can occur, one has to compare the rates of both processes.
We do so in the long-time limit of the coarsening process, i.e., for large peak masses and a large peak separation.
Using ${M \approx\Bar{\rho}\Lambda}$, we have from Eqs.~\eqref{eq:mKS-sigma-comp} and the coalescence rate obtained in Ref.~\cite{Kang.etal2007} [Eq.~(2.20) for $\Lambda\gg 1$; see also Ref.~\cite{Weyer.etalinpreparationa}] in the limit of large pattern (``coarsening limit'') wavelengths\footnote{
In the coarsening limit ${\Lambda\to\infty}$, peak competition will always be diffusion-limited because the mass transport over distances $\Lambda$ becomes slow.}
\begin{equation} 
    \sigma^- 
    = 
    8 \exp\left(-\frac{\Lambda}{\sqrt{D_c}}\right), 
    \qquad 
    \sigma^+ 
    = 
    \frac{T^2\bar\rho^2 \Lambda}{D_\rho D_c^{3/2}} \exp\left(-\frac{T\bar\rho}{2 D_\rho} \frac{\Lambda}{\sqrt{D_c}}\right) 
    \, .
\end{equation}
As the exponents in both growth rates differ by a factor of $\frac{T\bar\rho}{2 D_\rho}$, one has for large $\Lambda$
\begin{align} 
\sigma^+ \gg \sigma^- \quad &\mathrm{if}\quad \frac{T\bar\rho}{2 D_\rho} < 1\, , \\
\sigma^- \gg \sigma^+ \quad &\mathrm{if}\quad \frac{T\bar\rho}{2 D_\rho} > 1\, ,
\end{align}
such that the dominant coarsening scenario is set by the initial condition via fixing $\frac{T\bar\rho}{2 D_\rho} \lessgtr 1$ [cf.\ main text Fig.~2(c,d)].
The slower process will not significantly contribute to the coarsening process of the pattern at late times.

To derive the scaling of $\langle\Lambda\rangle(t)$ with time $t$ from the (dominant) rate $\sigma(M, \Lambda)$, we use ${\langle M\rangle \approx  \bar\rho \langle \Lambda\rangle}$ as ${\rho_- \approx 0}$. 
Furthermore, as is the central hypothesis in coarsening systems, we assume the mean peak mass $\bar\rho\langle \Lambda\rangle$ to be the only characteristic scale such that the mean collapse timescales must scale with ${\sigma(\mu \bar\rho\langle \Lambda\rangle, \nu \langle \Lambda\rangle)^{-1}}$, where $\mu$ and $\nu$ are scaling amplitudes~\cite{Weyer.etal2023,Bray2002}. 
As both timescales $(\sigma^+)^{-1}$ and $(\sigma^-)^{-1}$ increase exponentially in the peak-to-peak distance $\Lambda$, the time $t$ until an average distance $\langle \Lambda\rangle(t)$ is reached is dominated by the timescale of the most recent collapse ${\sigma(\mu \bar\rho\langle \Lambda\rangle(t), \nu \langle \Lambda\rangle(t))^{-1}}$~\cite{Weyer.etal2023}. 
Thus, we expect
\begin{equation}\label{eq:coarsening-scaling-1}
    t^{-1} 
    \sim 
    \sigma(\mu \bar\rho \langle \Lambda\rangle(t), \nu \langle \Lambda\rangle(t))
    \, .
\end{equation}
Using the faster one of the rates $\sigma^+$ and $\sigma^-$ as the mean collapse rate $\sigma$ in Eq.~\eqref{eq:coarsening-scaling-1}, the average length scale $\langle \Lambda\rangle$ thus scales with time like 
\begin{equation} 
    \langle\Lambda \rangle 
    \sim 
    \nu^{-1} \sqrt{D_c} \frac{\ln(t)}{\min\left(\frac{T\bar\rho}{2D_\rho}, 1\right)} + \mathcal{O}(\log\langle\Lambda\rangle)
    \, .
\end{equation}
Here, we  neglected logarithmic corrections ${\sim \log\langle\Lambda\rangle}$.
We verify this dependence of the logarithmic coarsening rate on the average cell density $\Bar{\rho}$ numerically in the main text, see Fig.~2(c,d).

\clearpage\newpage

\section{Weakly broken mass conservation -- effects of cell growth and death}
\label{sec:weakly-broken-mc}

In this section, we discuss plateau splitting and interrupted coarsening that is induced by accounting for cell growth and death, i.e., by breaking the conservation law for the cell density through a source term $\varepsilon \, s(\rho)$.
We consider the limit of weakly broken mass conservation where ${\varepsilon \ll 1}$.
This allows for a perturbative treatment of the effects of cell growth and death, using the pattern profiles constructed for the mass-conserving systems (see Sec.~\ref{sec:mass-comp-mKS}) as lowest-order approximation to the pattern profiles under the influence of cell growth and death.

As discussed in the main text, we assume that the growth and death term induces net cell growth at low and net cell death at high densities, as for example in logistic growth ${s(\rho) = \rho\, (p-\rho)}$ or for constant growth and linear death ${s(\rho) = p-\rho}$.
In particular, the stability of elementary mesa patterns requires ${s(\rho_-) > 0}$ and ${s(\rho_+) < 0}$, as discussed for two-component reaction--diffusion systems in Refs.~\cite{Brauns.etal2021,Weyer.etal2023}.

For concreteness, we consider peak patterns and the suppression of the mass-competition instability in the peak-coalescence scenario.
This is the case relevant to the mKS model at large peak mass.
The calculations are straight-forwardly extended to the other scenarios of peak competition and mass competition in mesa patterns as well (cf.\ Refs.~\cite{Brauns.etal2021,Weyer.etal2023} concerning these effects in mass-conserving reaction--diffusion systems).
In the following, we extend the methods introduced in Ref.~\cite{Brauns.etal2021} for two-component mass-conserving reaction–diffusion systems to address both plateau splitting and interrupted coarsening in chemotaxis-induced phase separation.

We denote the elementary stationary pattern, accounting for the effects of cell growth and death, by $\rhoStat^\varepsilon(x)$, $\etaStat^\varepsilon(x)$, and $\cStat^\varepsilon(x)$.
Because mass conservation is broken, the stationary mass-redistribution potential will no longer remain constant.
The profile $[\rhoStat^\varepsilon (x),\etaStat^\varepsilon (x)]$ is determined perturbatively around the stationary pattern of the corresponding mass-conserving system, i.e., one has
\begin{equation}
    [\rhoStat^\varepsilon (x),\etaStat^\varepsilon (x)] 
    = 
    [\rhoStat (x),\etaStat]
    +
    \varepsilon [\delta\rho_\varepsilon (x),\delta\eta_\varepsilon(x)] \, .
\end{equation}
where $[\delta\rho_\varepsilon,\delta\eta_\varepsilon]$ are characterizing the deviations from the corresponding mass-conserving pattern.
Importantly, we discuss below in Sec.~\ref{sec:plateau-splitting} that one has  ${[\delta\rho_\varepsilon,\delta\eta_\varepsilon] \sim {\cal O} (1)}$ for sufficiently short pattern wavelengths.
However,  one has ${\varepsilon[\delta\rho_\varepsilon,\delta\eta_\varepsilon] \sim {\cal O} (1)}$ if the pattern wavelength becomes large, that is, ${\Lambda \sim \varepsilon^{-1/2}}$.
In the limit ${\varepsilon\to0}$ coarsening is interrupted at much shorter wavelengths such that the threshold of interrupted coarsening $\Lambda_\mathrm{stop}$ can be obtained by assuming ${[\delta\rho_\varepsilon,\delta\eta_\varepsilon] \sim {\cal O} (1)}$ (up to logarithmic corrections) while $\varepsilon[\delta\rho_\varepsilon,\delta\eta_\varepsilon] \sim {\cal O} (1)$ is important for the threshold of plateau splitting $\Lambda_\mathrm{split}$.\footnote{
Mass competition weakens rapidly as the wavelength $\Lambda$ increases.
For example, in the mKS model the mass-competition rates decrease exponentially with the wavelength $\Lambda$ (see Sec.~\ref{sec:mass-comp-mKS}).
The same exponential decrease holds for general mesa patterns \cite{Weyer.etalinpreparationa}.
This gives a threshold of interrupted coarsening ${\Lambda_\mathrm{stop} \sim \log(1/\varepsilon)}$ (cf.\ Sec.~\ref{sec:interrupted-coarsening}).
In contrast, deformations of the pattern profile first arise in the plateaus and small deviations scale as $\sim \varepsilon\Lambda^2$ [cf.\ the continuity equation Eq.~(2) in the main text and Sec.~\ref{sec:plateau-splitting}].
Thus, deformations of the pattern (and splitting) are only relevant for ${\Lambda \sim \varepsilon^{-1/2}}$, which is much larger than $\Lambda_\mathrm{stop}$ in the limit ${\varepsilon \to 0}$.
}

Given that a family of stationary patterns $[\rhoStat (x),\etaStat]$ exists, depending on the average cell density $\Bar{\rho}_\mathrm{mc}$, what is the correct pattern to use as lowest-order approximation of $[\rhoStat^\varepsilon(x),\etaStat^\varepsilon(x)]$?
To answer this question, one must determine the average density $\bar{\rho}_\mathrm{mc}$ for that stationary pattern $[\rhoStat(x),\etaStat]$, which serves as the leading-order approximation for the modified profile $[\rhoStat^\varepsilon(x),\etaStat^\varepsilon(x)]$.
Interestingly, it is not necessarily the stationary pattern $[\rhoStat(x),\etaStat]$ with the same average cell density, ${\bar{\rho}_\mathrm{mc}=\bar{\rho}_\varepsilon}$, as the average density $\Bar{\rho}_\varepsilon = (1/L)\int_0^L\mathrm{d}x\, \rhoStat^\varepsilon(x)$ of $[\rhoStat^\varepsilon(x),\etaStat^\varepsilon(x)]$.
Rather, since mass conservation is broken in the system with cell growth and death, the average cell density is no longer an externally fixed control parameter but dynamically set by the balance of cell growth and death.
This condition of source balance, which then determines the stationary average density $\bar{\rho}_\varepsilon$, is obtained by integrating the modified continuity equation [Eq.~(2) in the main text] and demanding stationarity ${\partial_t \rho =0}$.
Integrating over the whole domain using no-flux or periodic boundary conditions then yields
\begin{equation}\label{eq:source-balance-approx}
    0 
    = \varepsilon \int_0^{\Lambda/2}\mathrm{d}x\, s(\rhoStat^\varepsilon) 
    = \varepsilon\int_0^{\Lambda/2}\mathrm{d}x\, s(\rhoStat) + \mathcal{O}(\varepsilon^2)\, ,
\end{equation}
in the case that one has $[\delta\rho_\varepsilon,\delta\eta_\varepsilon] \sim {\cal O} (1)$.
Thus, the mass-conserving stationary pattern approximating the profile $[\rhoStat^\varepsilon,\etaStat^\varepsilon]$ to lowest order fulfills
\begin{equation}
    0 
    = \int_0^{\Lambda/2}\mathrm{d}x\, s(\rhoStat)
    \, .
\end{equation}
Because $\rhoStat$ is a function of its average density $\Bar{\rho}_\mathrm{mc}$, or equivalently of $\etaStat$ or the peak mass, its solution fixes the average density $\Bar{\rho}_\mathrm{mc}$ of the profile $[\rhoStat (x),\etaStat]$ of the mass-conserving system that gives the leading-order approximation of $[\rhoStat^\varepsilon,\etaStat^\varepsilon]$.
Below, we determine the source-balance condition explicitly for the mKS model with a logistic source term ${s(\rho) = \rho \, (p-\rho)}$, resulting in Eq.~\eqref{eq:peak-mass-from-sourceBalance-mKS} for the peak mass in dependence of the production rate $p$.

As we discussed above, the approximation Eq.~\eqref{eq:source-balance-approx} is strictly valid only for pattern wavelengths $\Lambda$ on which the pattern profile does not change significantly compared to the mass-conserving pattern, that is, the approximation only holds far below the threshold for plateau splitting (see Sec.~\ref{sec:plateau-splitting}).
Before addressing interrupted coarsening, we will first discuss plateau splitting, in which case the deviations of the pattern profile from the mass-conserving profile are crucial.

\subsection{Plateau splitting}
\label{sec:plateau-splitting}

In the limit of weak source strength, ${\varepsilon \ll 1}$, cell growth and death will induce only weak gradients affecting the pattern primarily on long length scales.
As shown below, the induced gradients are of the order $\mathcal{O}(\sqrt{\varepsilon})$, that is, significant variations occur on length scales ${\sim \varepsilon^{-1/2}}$.
Thus, the amplitudes $\varepsilon \delta\rho_\varepsilon$ and $\varepsilon \delta\eta_\varepsilon$, characterizing the deviations from the corresponding mass-conserving pattern, grow on sufficiently long distances $\mathcal{O}(\varepsilon^{-1/2})$, and can become of the order $\mathcal{O}(1)$ far away from the interface and peak regions.
In contrast, the sharp peaks and interfaces confined to narrow peak or interface regions of widths ${\sim\ell_\mathrm{int}\ll \varepsilon^{-1/2}}$ remain unchanged to lowest order. 
The significant deviations in the pattern plateaus then trigger the splitting of plateaus with a length ${\sim\varepsilon^{-1/2}}$, as we will discuss in the following.

Now, what is the deformation in the low-density plateau between two peaks induced by cell growth and death?
Assuming that the cells show a net growth at low and net death at high densities, net cell growth in the low-density plateau increases the cell density there [cf.\ Fig.~3(b) in the main text], which in turn alters the chemoattractant concentration as prescribed by the reaction term $f$.
Together, these changes in cell and chemoattractant density induce a profile in the mass-redistribution potential.
The gradients of this profile must ensure that the combined diffusive and chemotactic mass transport of cells from the plateau to the degradation region at high cell densities in the peak [cf.\ Fig.~3(a) in the main text] balances cell growth in the plateau.
This latter requirement follows from the modified continuity equation Eq.~(2) in the main text at stationarity $\partial_t\rho =0$.

Therefore, to determine an explicit expression for the stationary profiles in the plateau regions, we start from the modified continuity equation [cf.\ Eq.~(2) in the main text] at stationarity ${\partial_t\rho = 0}$.
Since the source term scales as ${s(\rho) \sim \mathcal{O}(1)}$, one can rewrite this equation by defining the rescaled profile ${[\tilde{\rho}_\mathrm{stat}^\varepsilon (x),\tilde{\eta}_\mathrm{stat}^\varepsilon (x)] \equiv [\rhoStat^\varepsilon(\sqrt{\varepsilon}x),\etaStat^\varepsilon(\sqrt{\varepsilon}x)]}$ as follows
\begin{equation}\label{eq:eps-plateau-profile-eq-rescaled}
    0 = T \partial_y
    \big[ \chi_\rho(\tilde{\rho}_\mathrm{stat}^\varepsilon) \, 
    \tilde{\rho}_\mathrm{stat}^\varepsilon \, 
    \partial_y \tilde{\eta}_\mathrm{stat}^\varepsilon
    \big] 
    + 
    s(\tilde{\rho}_\mathrm{stat}^\varepsilon)
    \, ,
\end{equation}
where ${y \equiv \sqrt{\varepsilon}x}$.
Here we have assumed that same scaling holds for $\tilde{\rho}_\mathrm{stat}^\varepsilon$ and $\tilde{\eta}_\mathrm{stat}^\varepsilon$, since they are linked by the reaction term $f$ which is strong [of order $\mathcal{O}(1)$] compared to the source term.
Thus, these rescaled profiles have source-induced gradients which are of the order $\mathcal{O}(1)$ in the plateau regions of the pattern.
Equation~\eqref{eq:eps-plateau-profile-eq-rescaled} therefore implies that the original profiles $[\rhoStat^\varepsilon(x),\etaStat^\varepsilon(x)]$ have gradients of the order $\mathcal{O}(\sqrt{\varepsilon})$.
Because one has ${[\rhoStat^\varepsilon(x),\etaStat^\varepsilon(x)] \approx [\rho_-,\etaStat] + \varepsilon \, [\delta\rho_\varepsilon(x), \delta\eta_\varepsilon(x)]}$ in the plateau regions, also the deviations $\varepsilon \, [\delta\rho_\varepsilon(x), \delta\eta_\varepsilon(x)]$ have gradients of the order $\mathcal{O}(\sqrt{\varepsilon})$, as we anticipated at the beginning of this section.

To solve for the profiles $\tilde{\rho}_\mathrm{stat}^\varepsilon$ and $\tilde{\eta}_\mathrm{stat}^\varepsilon$, we now consider the equation of motion for the mass-redistribution potential, Eq.~\eqref{eq:KS-eta}.
Inserting the rescaled profiles into Eq.~\eqref{eq:KS-eta} at stationarity $\partial_t\eta=0$, one obtains
\begin{equation}\label{eq:nullcline-restriction-plateaus}
    0 = -\chi_c(\tilde{c}_\mathrm{stat}^\varepsilon)\Tilde{\Tilde{f}}(\tilde{\rho}_\mathrm{stat}^\varepsilon, \tilde{\eta}_\mathrm{stat}^\varepsilon) + \mathcal{O}(\varepsilon)\,.
\end{equation}
Here we have used that all gradient terms in Eq.~\eqref{eq:KS-eta} can be neglected because the source term only introduces weak gradients ${\sim\mathcal{O}(\sqrt{\varepsilon})}$.
As a result, Eq.~\eqref{eq:nullcline-restriction-plateaus} show that, to lowest order in $\varepsilon$, the plateau profile follows the nullcline ${\tilde{\tilde{f}} = 0}$, i.e., to fulfill Eq.~\eqref{eq:nullcline-restriction-plateaus} the plateau profiles must fulfill ${\etaStat^\varepsilon \approx \eta^*(\rhoStat^\varepsilon)}$ (cf.\ Ref.~\cite{Brauns.etal2021}). 
The stationary plateau profile $\rhoStat^\varepsilon (x)$ is then determined by the stationary modified continuity equation Eq.~\eqref{eq:eps-plateau-profile-eq-rescaled}, which yields
\begin{equation}\label{eq:eps-plateau-profile-eq}
    0 \approx T \partial_x\left[ \chi_\rho(\rhoStat^\varepsilon)\rhoStat^\varepsilon \partial_x \eta^*(\rhoStat^\varepsilon)\right] + \varepsilon s(\rhoStat^\varepsilon)\, .
\end{equation}
Here, we have dropped the rescaling and wrote the profile equation in terms of the original field $\rhoStat^\varepsilon$.

Within the sharp-interface (sharp-peak) approximation, the plateau profile solving this equation must be matched to the solution in the peak or interface region.
Because the pattern remains unaffected in the peak or interface region at ${x \approx x_\mathrm{peak}}$ and the stationary peak profile approaches the plateau density exponentially just outside the peak, ${\rhoStat \to \rho_-}$ (cf.\ Ref.~\cite{Weyer.etalinpreparationa}), the plateau-profile equation Eq.~\eqref{eq:eps-plateau-profile-eq} must be solved using the boundary condition
\begin{equation}
    \rhoStat^\varepsilon(x_\mathrm{peak})
    \approx 
    \rho_-
    \,.
\end{equation}
For concreteness, let us now consider the plateau between two equally-sized peaks located at ${x = 0}$ and ${x = \Lambda}$.
Then, the plateau profile is symmetric around $\Lambda/2$, and we can again consider only the elementary (half-peak) pattern on the domain $[0,\Lambda/2]$ with no-flux boundary conditions.
Integrating the plateau-profile equation, Eq.~\eqref{eq:eps-plateau-profile-eq}, over a part of the domain $[y,\Lambda/2]$ shows that the gradient $\partial_x\etaStat^\varepsilon$ is positive if ${s(\rho_-) > 0}$, that is, for net cell production in the low-density plateau.
Because the lateral stability of the plateau (cf.\ Sec.~\ref{sec:plateau-stability}) ensures ${\partial_\rho\eta^* > 0}$ such that ${\partial_x\rhoStat^\varepsilon > 0}$ holds as well.
Thus, the plateau density rises between the peaks and the plateau profile has a maximum at $\Lambda/2$.
When the maximal plateau density surpasses the density $\rho_\mathrm{max}$ at which the nullcline $\eta^*(\rho)$ reaches its maximum, the nullcline slope $\partial_\rho\eta^*$ becomes negative, signifying that the maximum of the plateau becomes laterally unstable [cf.\ Eq.~(4) in the main text].
As the result of this lateral instability, a new peak is nucleated that splits the plateau in half (cf.\ Refs.~\cite{Kolokolnikov.etal2007,Brauns.etal2021}).

We now explicitly calculate the plateau profile and the threshold of plateau splitting, $\Lambda_\mathrm{split}$, for the mKS model with the logistic source term ${s(\rho) = \rho (p-\rho)}$.
This model has the nullcline 
(cf.\ Sec.~\ref{sec:reactive-area-balance})
\begin{equation}
    \eta^*(\rho) = \frac{D_\rho}{T}\log(\rho) -\rho\,.
\end{equation}
For sufficiently large average densities $\Bar{\rho}$, the density $\rho_-$ of the low-density plateau is small such that ${|\log(\rho_-)| \gg \rho_-}$ holds.\footnote{
One has ${\rho_-\to0}$ for large average densities $\Bar{\rho}$, that is for large peaks because the density of the low-density plateau fulfills $\rho_-\approx\exp[T\etaStat/D_\rho]$ (FBS-NC intersection) and $\etaStat\sim - M$ as $M\to\infty$ (cf.\ Sec.~\ref{sec:mass-comp-mKS}).
Also note that we use densities with rescaled units such that $T\rho/D_\rho$ and $T\eta/D_\rho$ are dimensionless.
}
Consequently, the nullcline can be approximated in the low-density plateau as 
\begin{equation}
    \eta^*(\rho) 
    \approx 
    \frac{D_\rho}{T}\log(\rho)
    \,.
\end{equation}
With this, the plateau-profile equation Eq.~\eqref{eq:eps-plateau-profile-eq} becomes 
\begin{equation}
    D_\rho \partial_x^2 \rhoStat^\varepsilon = - \varepsilon \rhoStat^\varepsilon (p - \rhoStat^\varepsilon) \approx -\varepsilon p \rhoStat^\varepsilon\, ,
\end{equation}
where the approximation of the source term holds at low densities.
Considering the elementary pattern on the domain ${I = [0,\Lambda/2]}$ with reflective boundary conditions and the peak positioned at the boundary ${x = 0}$,
the approximate plateau-profile equation (a Helmholtz equation) with the boundary conditions ${\rhoStat^\varepsilon(0) = \rho_-}$ and ${\partial_x\rhoStat^\varepsilon(\Lambda/2)=0}$ is solved by
\begin{equation}\label{eq:mKS-splitting-plateau-profile}
    \rhoStat^\varepsilon(x) \approx \frac{\rho_-}{\cos\left[\sqrt{\frac{\varepsilon p}{D_\rho}} \frac{\Lambda}{2}\right]} \cos\left[\sqrt{\frac{\varepsilon p}{D_\rho}} \left(\frac{\Lambda}{2}-x\right)\right] .
\end{equation}
This result agrees with proposition 4.1 of Ref.~\cite{Kolokolnikov.etal2014}.
We can correct for a finite peak width by replacing $\Lambda/2$ by (a heuristic estimate of) the half-length $L_-$ of the low-density plateau in the denominator 
of the prefactor.

This profile has a maximum at $\Lambda/2$.
For sufficiently large peaks, the plateau density approximately vanishes, ${\rho_- \approx 0}$ (see footnote above).
Because $\rho_-$ appears in the prefactor of the plateau profile, the amplitude of the plateau profile becomes significant only if the denominator of the prefactor in Eq.~\eqref{eq:mKS-splitting-plateau-profile} also becomes small.
Thus, the cosine must become zero (approximately), showing that the plateau density only significantly grows if one has
\begin{equation}\label{eq:splitting-denominator-zero}
    \sqrt{\frac{\varepsilon p}{D_\rho}} L_- 
    \approx
    \frac{\pi}{2}
    \,.
\end{equation}
The maximum plateau density diverges at the equality.
As a result, in the limit ${\rho_- \to 0}$, that is, for large peaks as discussed above, the condition Eq.~\eqref{eq:splitting-denominator-zero} gives the plateau length $L_-$ at which the maximum density crosses the density of the nullcline maximum $\rho_\mathrm{max}$ and enters the laterally unstable density regime.
Defining the interface or half-peak width ${\ell_\mathrm{int} = \Lambda/2-L_-}$, the threshold wavelength of plateau splitting $\Lambda_\mathrm{split}$ is
\begin{equation}\label{eq:mKS-Lambda-split}
     \frac{\Lambda_\mathrm{split}}{2} = \ell_\mathrm{int} + \frac{\pi}{2} \sqrt{\frac{D_\rho}{\varepsilon p}}
     \, .
\end{equation}
In the sharp-peak limit, the interface width $\ell_\mathrm{int}$ is negligible.
For small source strength $\varepsilon$, the splitting wavelength thus scales as
\begin{equation}
    \Lambda_\mathrm{split}(\varepsilon) \sim (\varepsilon p)^{-1/2}.
\end{equation}
Rescaling the space variable $x$ in the plateau-profile equation Eq.~\eqref{eq:eps-plateau-profile-eq} using ${y = \sqrt{\varepsilon} x}$ shows that the scaling ${\Lambda_\mathrm{split}(\varepsilon) \sim \varepsilon^{-1/2}}$ holds in general.
The scaling is independent of the plateau density $\rho_-$ because the growth term $\sim \varepsilon p\rho$ grows as the lower plateau bends upwards.
As we will see below, this is different for interrupted coarsening, for which $\varepsilon \rho_-$ is the relevant strength of the source term far below the splitting threshold.

The threshold $\Lambda_\mathrm{split}$ determined by Eq.~\eqref{eq:mKS-Lambda-split} is compared to simulations in the main text in Fig.~3(c), estimating the interface width by ${\ell_\mathrm{int} = 2\sqrt{D_c}}$ [cf.\ $\ell_\mathrm{int}$ defined in Ref.~\cite{Weyer.etalinpreparationa}].

\subsection{Interrupted coarsening}
\label{sec:interrupted-coarsening}

We now discuss interrupted coarsening.
We do so for the coalescence scenario and consider, as in Sec.~\ref{sec:mass-comp-QSS}, a peak centered at ${x = 0}$ on the domain ${I = [-\Lambda/2,\Lambda/2]}$ with reflective boundaries [see Figs.~2(b),~3(a) in the main text].

We discussed in the main text that interrupted coarsening occurs because the source term counteracts the mass-competition instability of the mass-conserving system.
While the mass-competition instability redistributes mass from the smaller to the larger peak or the longer to the shorter plateau [cf.\ Figs.~2(a,b)], the source term leads to net cell growth/production at the smaller peak or longer plateau, counteracting the mass transport due to the mass-competition instability.
The reason is that we assume net cell growth at low densities $\rho$ and net cell death at large cell densities.
The coarsening process is interrupted at the threshold wavelength $\Lambda_\mathrm{stop}(\varepsilon)$, where the (largest) growth rate of the mass-competition instability, affected by cell growth and death, reaches zero and turns negative.
We determine the growth rate in the diffusion-limited regime following the intuitive argument given in Ref.~\cite{Brauns.etal2021}.
Similar to the strictly mass-conserving case, we assume that peak coalescence is slow compared to the relaxation of the stationary elementary profile (cf.\ Ref.~\cite{Weyer.etalinpreparationa}).
It was shown for two-component reaction--diffusion systems with weakly broken mass conservation using a singular perturbation analysis that the threshold of interrupted coarsening $\Lambda_\mathrm{stop}$ determined in the diffusion-limited regime is exact also in the reaction-limited regime of mass competition \cite{Weyer.etal2023}.
Reaction limitation only gives rise to an overall prefactor of the rate in these reaction--diffusion systems.
We expect the same can be shown for Keller--Segel systems by generalizing the singular perturbation analysis for the strictly mass-conserving system given in Ref.~\cite{Weyer.etalinpreparationa}.

To assess the stability of the centered position ${x = 0}$, we consider a small perturbation $[\delta\rho(x),\delta\eta(x)]$ of the stationary pattern $[\rhoStat^\varepsilon,\etaStat^\varepsilon]$ that translates the peak to ${x = \delta x}$.
To determine whether the peak moves back to ${x = 0}$ or continues moving toward the boundary at ${x = \Lambda/2}$,
we integrate the modified continuity equation [Eq.~(2) in the main text] over the interval $[\delta x, \Lambda/2]$ on the right side of the peak.
This gives the mass change
\begin{equation}\label{eq:IC-mass-change}
    \frac12 \partial_t \delta M 
    = - T \chi_\rho(\hat{\rho}) \, \hat{\rho} \, \partial_x\delta\eta|_{x=\delta x} 
    + 
    \varepsilon \int_{\delta x}^{\Lambda/2}\mathrm{d}x\, s(\rho)
    \, .
\end{equation}
Given that changes in the plateau region are exponentially small in the domain length $\Lambda/2$ or of order $\varepsilon$ for domain lengths ${\Lambda \ll \Lambda_\mathrm{split}}$ (see Ref.~\cite{Weyer.etalinpreparationa} and Sec.~\ref{sec:plateau-splitting}), the mass change $\delta M/2$ is mainly due to the translation of the peak, that is, one has
\begin{equation}
    \frac12 \partial_t \delta M  
    \approx 
    (\hat{\rho}-\rho_-) \, 
    \partial_t \delta x
    \, .
\end{equation}
The first term on the right-hand side of Eq.~\eqref{eq:IC-mass-change} describes the redistribution of mass through the peak due to the gradients in the mass-redistribution potential induced by the peak translation.
Following the Supplemental Material, Sec.~7.2 of Ref.~\cite{Brauns.etal2021}, we approximate this using the (diffusion-limited) mass-competition rate in the coalescence scenario $\sigma_\mathrm{D}^-$ of the strictly mass-conserving system, giving
\begin{equation}\label{eq:PS-flux-approx}
    - T\chi_\rho(\hat{\rho}) \, 
    \hat{\rho}\, \partial_x\delta\eta|_{x=\delta x} 
    \approx 
    \sigma_\mathrm{D}^- \, 
    \big[ 1 + \mathcal{O}(\varepsilon) \big] \, \frac{\delta M}{2}\, ,
\end{equation}
because $\partial_t \delta M/2 = \sigma_\mathrm{D}^- \delta M/2$ holds in the mass-conserving case (cf.\ Sec.~\ref{sec:mass-comp-QSS}).

The second term in Eq.~\eqref{eq:IC-mass-change} is the contribution due to the source term.
Anticipating that, in the linear regime for small $\delta M$ or equivalently small $\delta x$, the source contribution will be linear in the deviation, we define the source-induced rate
\begin{equation}
    -\sigma_\mathrm{S} \frac{\delta M}{2} 
    \equiv 
    \int_{\delta x}^{\Lambda/2}\mathrm{d}x\, s(\rho)
    \, .
\end{equation}
Setting the minus sign, we anticipate the stabilizing effect of the source term.
To calculate the source contribution, we use that the source term must be balanced at a stationary pattern [cf.\ Eq.~\eqref{eq:source-balance-approx}], i.e., one has
${0 = \int_0^{\Lambda/2}\mathrm{d}x\, s(\rhoStat^\varepsilon)}$. 
Adding this vanishing contribution to the source contribution, one obtains
\begin{align}
   \int_{\delta x}^{\Lambda/2}\mathrm{d}x\, s(\rho) &= \int_{\delta x}^{\Lambda/2}\mathrm{d}x\, s(\rho) - \int_{\delta x}^{\Lambda/2+\delta x}\mathrm{d}x\, s(\rhoStat^\varepsilon(x-\delta x))\label{eq:IC-source-integral-exact}\\
    &\approx - \delta x\,  s(\rhoStat^\varepsilon(\Lambda/2)) + \int_{\delta x}^{\Lambda/2}\mathrm{d}x\, [ s(\rho) -s(\rhoStat^\varepsilon(x-\delta x))]\, .\label{eq:IC-source-integral}
\end{align}
Because we assume mass competition to be slow compared to the relaxation of the peak profile onto the stationary profile, the decomposition Eq.~\eqref{eq:IC-source-integral} shows that the source contribution to the growth rate $\sigma_\mathrm{S}$ is only due to changes in the plateaus.
Specifically, the first term is evaluated in the plateau and the second integral vanishes in the peak region because there the profile of the shifted peak fulfills ${\rho(x)\approx\rhoStat^\varepsilon(x-\delta x)}$.

Consequently, if the plateaus are short compared to the length at which they split (cf.\ Sec.~\ref{sec:plateau-splitting}), the source term only induces small changes in the plateau densities,i.e., it holds  ${\rhoStat^\varepsilon= \rho_-+\mathcal{O}(\varepsilon)}$, the integral on the right-hand side in Eq.~\eqref{eq:IC-source-integral} is of the order $\mathcal{O}(\varepsilon)$ as well.
One then finds by combining Eqs.~\eqref{eq:IC-mass-change}--\eqref{eq:IC-source-integral}
\begin{equation}
    \partial_t \delta x \approx \left[ \sigma_\mathrm{D}^- - \varepsilon\frac{s(\rho_-)}{\hat{\rho}-\rho_-} \right] \delta x\, ,
\end{equation}
that is, the mass-competition growth rate under the influence of the source term in the coalescence scenario is
\begin{equation}\label{eq:eps-sigma}
    \sigma_\mathrm{D,-}^\varepsilon(\Lambda) \approx \sigma_\mathrm{D}^-(\Lambda) - \varepsilon \sigma_\mathrm{S}\, ,
\end{equation}
with $\sigma_\mathrm{S} \approx s(\rho_-)/(\hat{\rho}-\rho_-)$, and denoted by $\sigma_\mathrm{-}^\varepsilon$ in the main text.
As argued in the main text, for $s(\rho_-)>0$, the source term counteracts the mass-competition instability of the strictly mass-conserving system.
The threshold $\Lambda_\mathrm{stop}$ of interrupted coarsening follows from solving
\begin{equation}
    \sigma_\mathrm{D,-}^\varepsilon(\Lambda_\mathrm{stop})=0\, .
\end{equation}
Evaluated below for the mKS system, this relation is shown as the dashed line in Fig.~3(c) in the main text.

\paragraph{Interrupted coarsening at wavelengths close to the plateau-splitting threshold.\;---} 
When the threshold of plateau splitting is approached, one has ${\rhoStat^\varepsilon- \rho_- = \mathcal{O}(1)}$ in the plateaus (cf.\ Sec.~\ref{sec:plateau-splitting}).
Therefore, one has to consider the source rate $\sigma_\mathrm{S}$ in more detail.
As discussed above, Eq.~\eqref{eq:IC-source-integral-exact} shows that the source contribution is only due to changes in the plateau region because the translated stationary peak profile approximates the peak region, that is, one has $\rho\approx\rhoStat^\varepsilon(x-\delta x)$ because we assume that the relaxation of the peak profile onto the stationary profile is fast compared to the coalescence process (QSS approximation).
Thus, the integral difference in Eq.~\eqref{eq:IC-source-integral-exact} can be calculated by comparing the plateau profiles of the shifted peak $\rho$ and the symmetric stationary pattern $\rhoStat^\varepsilon$.
Because we assume peak coalescence to be slow, the plateau profile of the shifted peak $\rho(x)$ is approximated by the stationary plateau profile, Eq.~\eqref{eq:eps-plateau-profile-eq}, which is induced by the source term and calculated above to determine the threshold of plateau splitting.
Importantly, the (stationary) plateau profile differs from the plateau profile of the symmetric stationary pattern $\rhoStat^\varepsilon(x)$ because the peak is shifted and, therefore, the right plateau is shorter (and the left one is longer).
Thus, we denote by $\rho_\mathrm{plateau}^\varepsilon$ the solution to Eq.~\eqref{eq:eps-plateau-profile-eq} for the boundary conditions ${\rho = \rho_-}$ at ${x = \ell_\mathrm{int}+\delta x}$ and a reflective boundary at ${x = \Lambda/2}$ with the (heuristic) approximation of the peak width $\ell_\mathrm{int}$ introduced for plateau splitting (see Sec.~\ref{sec:plateau-splitting}), and which is set to zero in the sharp-peak approximation.
This profile describes the (right) plateau of the shifted peak $\rho(x)$ on the domain $I$ that is shortened by $\delta x$.
Moreover, we denote by $\rho_{\mathrm{plateau},0}^\varepsilon$ the solution for the boundary conditions ${\rho = \rho_-}$ at ${x = \ell_\mathrm{int}}$ and a no-flux condition at ${x = \Lambda/2}$, which gives the plateau profile of the unshifted peak $\rhoStat^\varepsilon(x)$.
With this, one can approximate [using Eq.~\eqref{eq:IC-source-integral-exact}]
\begin{align}
    \int_{\delta x}^{\Lambda/2}\mathrm{d}x\, s(\rho) &= \int_{\delta x}^{\Lambda/2}\mathrm{d}x\, s(\rho) - \int_0^{\Lambda/2}\mathrm{d}x\, s(\rhoStat^\varepsilon(x))\\
    &\approx \int_{\ell_\mathrm{int}+\delta x}^{\Lambda/2}\mathrm{d}x\, s(\rho_\mathrm{plateau}^\varepsilon) - \int_{\ell_\mathrm{int}}^{\Lambda/2}\mathrm{d}x\, s(\rho_{\mathrm{plateau},0}^\varepsilon)\\
    &\equiv -(\partial_{L_-} S_\mathrm{plateau}) \delta x + \mathcal{O}(\delta x^2)\, .
    \label{eq:eps-sigma-PS}
\end{align}
For the approximation in the second line we use that the peak region does not contribute to the integral difference as discussed above.
In the third line, we define the source contribution of a plateau of half-length $L_-$
\begin{equation}\label{eq:source-contribution-plateau}
    S_\mathrm{plateau}(L_-) \equiv \int_{\Lambda/2-L_-}^{\Lambda/2}\mathrm{d}x\, s(\rho_\mathrm{plateau}^\varepsilon(x;L_-)) = \int_0^{L_-}\mathrm{d}\tilde{x}\, s(\rho_\mathrm{plateau}^\varepsilon(\Lambda/2-\tilde{x};L_-)),
\end{equation}
using ${\tilde{x} = \Lambda/2-x}$ and the plateau-profile solution $\rho_\mathrm{plateau}^\varepsilon(x;L_-)$ to Eq.~\eqref{eq:eps-plateau-profile-eq} with the boundary condition ${\rho = \rho_-}$ at ${x = \Lambda/2-L_-}$ and a reflective boundary at ${x = \Lambda/2}$.
The third line then gives the first-order change of this contribution for a plateau shortened by a small amount $\delta x$ around ${L_-=\Lambda/2-\ell_\mathrm{int}}$.

Employing Eq.~\eqref{eq:eps-sigma-PS} and Eq.~\eqref{eq:IC-mass-change}, one more generally has, not only for ${\Lambda \ll \Lambda_\mathrm{split}}$, the growth rate of peak coalescence under the influence of cell growth and death
\begin{equation}\label{eq:IC-rate-closePS}
    \sigma_\mathrm{D,-}^\varepsilon(\Lambda) \approx \sigma_\mathrm{D}^-(\Lambda) - \varepsilon\frac{\partial_{L_-}S_\mathrm{plateau}}{\hat{\rho}-\rho_-}\, ,
\end{equation}
with ${\sigma_\mathrm{S} \approx \partial_{L_-}S_\mathrm{plateau}/(\hat{\rho}-\rho_-)}$.
Note that we haven't shown this to be a systematic expansion because we assume $\sigma_\mathrm{D}^-(\Lambda)$ to still be a valid approximation of the mass transport through the peak although the plateau densities have changed by $\mathcal{O}(1)$ compared to the mass-conserving system.
One can improve the approximation by correcting the peak mass $M$ by the plateau contribution to the source balance integral Eq.~\eqref{eq:source-balance-approx} (cf.\ Ref.~\cite{Kolokolnikov.etal2014}).
However, for a systematic approximation, one also has to analyze how the exponential tails emanating from the peak change given that $\rho_-$ is spatially varying.
Nonetheless, setting Eq.~\eqref{eq:IC-rate-closePS} to zero gives the solid line in Fig.~3(c) in the main text for the mKS system with logistic growth, which agrees remarkably well with the numerical results for the interrupted-coarsening threshold $\Lambda_\mathrm{stop}$. 

\smallskip

\paragraph{Explicit evaluation for the mKS model with logistic growth.\;---}
As the last step, we evaluate the two rate expressions Eqs.~\eqref{eq:eps-sigma},~\eqref{eq:eps-sigma-PS} for the mKS model with the logistic source term $s(\rho) = \rho (p-\rho)$.
The stationary profile $\rhoStat$ of the mass-conserving system that approximates the profile in the system including cell growth and death, is fixed by Eq.~\eqref{eq:source-balance-approx}.
Because one has $s(\rho)\approx 0$ in the plateaus for logistic growth, the peak mass is selected by the balance of growth and death in the low- and high-density regions of the peak itself while the plateaus do not contribute significantly.
Hence, the selected peak mass $M$ is independent of the plateau length and the condition Eq.~\eqref{eq:source-balance-approx} can be solved by integrating over the infinite line and using the peak profile $\rhoStat^\infty$ Eq.~\eqref{eq:mKS-stat-rho}, which gives
\begin{equation}
    0 
    \approx 
    \int_{-\infty}^\infty\mathrm{d}x\, \rhoStat^\infty = p M - \frac{M^3 T}{12 D_\rho D_c}
    \,.
\end{equation}
Thus, the peak mass is fixed to
\begin{equation}\label{eq:peak-mass-from-sourceBalance-mKS}
    M 
    = 
    \sqrt{\frac{12 p D_\rho D_c}{T}}
    \, ,
\end{equation}
which has to be inserted in all the following formulas if not done explicitly.

The diffusion-limited mass-competition rate for the coalescence scenario in the strictly mass-conserving system is [cf.\ Ref.~\cite{Weyer.etalinpreparationa} and Eq.~(2.20) for ${\Lambda \gg 1}$ of Ref.~\cite{Kang.etal2007}]
\begin{equation}
    \sigma_\mathrm{D}^- \approx 2 \frac{T M}{D_c^{3/2}}\exp\left( -\frac{\Lambda}{\sqrt{D_c}}\right).
\end{equation}
Moreover, with $s(\rho_-) \approx p \rho_-$ one has
\begin{align}
    \varepsilon p \frac{\rho_-}{\hat{\rho}-\rho_-} &\approx \varepsilon p \frac{\rho_-}{\hat{\rho}} \approx \varepsilon p \frac{\exp\left(\frac{T}{D_\rho}\etaStat\right)}{\hat{\rho}} \approx 4\varepsilon p \exp\left(-\frac{T M}{2 D_\rho \sqrt{D_c}}\right).\label{eq:mKS-source-contr}
\end{align}
Thus, far below the threshold of plateau splitting, coarsening is counteracted by a source term of strength $\varepsilon p \rho_-$.
Because the density of the lower plateau $\rho_-\approx 0$ is small for large peak masses, the source term appears effectively much weaker for counteracting coarsening, i.e., for determining $\Lambda_\mathrm{stop}$, than the source strength $\varepsilon p$ relevant in determining the threshold of plateau splitting $\Lambda_\mathrm{split}$.
For comparison, if the linear reaction term $s(\rho) = p-\rho$ is chosen, both thresholds would depend on $\varepsilon p$.
Thus, the threshold wavelength $\Lambda_\mathrm{stop}$ is larger compared to the threshold $\Lambda_\mathrm{split}$ in the case of logistic growth than for a linear source term.

As the two thresholds approach each other with increasing source strength $\varepsilon$ [see Fig.~3(c) in the main text], one should use Eq.~\eqref{eq:eps-sigma-PS} to determine the threshold of interrupted coarsening. To calculate $\partial_{L_-}S_\mathrm{plateau}$, we insert the plateau profile of the mKS system with a logistic source term Eq.~\eqref{eq:mKS-splitting-plateau-profile} in Eq.~\eqref{eq:source-contribution-plateau} which gives
\begin{equation}
    S_\mathrm{plateau}(L_-) = \int_0^{L_-}\mathrm{d}x\, \frac{\rho_-}{\cos\left[\sqrt{\frac{\varepsilon p}{D_\rho}} L_-\right]} \cos\left[\sqrt{\frac{\varepsilon p}{D_\rho}} x\right] = \sqrt{\frac{D_\rho}{\varepsilon p}}\rho_- \tan\left[\sqrt{\frac{\varepsilon p}{D_\rho}} L_-\right].
\end{equation}
From this, one obtains 
\begin{equation}
    \partial_{L_-} S_\mathrm{plateau} 
    \approx 
    \frac{\rho_-}{ \cos^2\left(\sqrt{\frac{\varepsilon p}{D_\rho}} L_-\right)}
    \, ,
\end{equation}
and the source contribution to the growth rate is corrected to [cf.\ Eq.~\eqref{eq:mKS-source-contr}]
\begin{equation}
    \sigma_\mathrm{S} \approx p \frac{\rho_-}{\hat{\rho}-\rho_-} \left[\cos\left(\sqrt{\frac{\varepsilon p}{D_\rho}} L_-\right)\right]^{-2}.
\end{equation}
This shows that the stabilizing source contribution in $\sigma_{\mathrm{D},-}^\varepsilon$ diverges as the threshold of plateau splitting $\Lambda_\mathrm{split}$ is approached [cf.\ Eq.~\eqref{eq:mKS-Lambda-split}], leading to the downward bending of $\Lambda_\mathrm{stop}$ seen in the solid line in Fig.~3(c) in the main text.
The reason for this divergence is that the density change of the plateau maximum with the plateau length diverges, which then leads to a diverging source contribution $\sigma_\mathrm{S} \sim \int_{\delta x}^{\Lambda/2}\mathrm{d}x\, s(\rho)$ with the shift $\delta x$.
Because the plateau-splitting threshold $\Lambda_\mathrm{split}$ corresponds to a fold bifurcation at which a new peak is inserted into the pattern (see Ref.~\cite{Kolokolnikov.etal2007,Brauns.etal2021}), the diverging rate of change of the density of the plateau maximum is generic.
Consequently, the downward-bend of $\Lambda_\mathrm{stop}$ is generic, and the two thresholds $\Lambda_\mathrm{stop}$ and $\Lambda_\mathrm{split}$ will not cross but only approach each other within our (uncontrolled)  approximation.

\clearpage\newpage

\section{Numerical simulation}
\label{sec:numerics}
Simulations of the mKS model are performed using the time-dependent finite-element solver of \textsc{Comsol} Multiphysics (Version 6.0) \cite{Comsol} on a regular one-dimensional grid with a grid-element size of 0.001 [cf.\ Fig.~2(c,d) in the main text]. 
The average length scale of one simulation, $\langle\Lambda\rangle(t) = L/N(t)$, is calculated by dividing the system length $L$ by the number of peaks $N(t)$.
The peaks are identified via the python library scipy's \texttt{signal.find\_peaks()} function \cite{Virtanen.etal2020}.
The length scales in Fig.~2(\textit{d}) of the main text are ensemble averages over multiple sweeps each, with the number of simulations indicated in the figure caption.
The analysis of the finite-element simulations is performed using Python.

The mKS model with a logistic growth term is also simulated using Comsol Multiphysics (Version 6.1) \cite{Comsol}.
The stationary patterns and their stability are determined using a finite-differences discretization of the model equations based on first- and second-order central differences.
The stationary patterns are zeros of these equations, which are found by a Newton iteration.
The family of stationary patterns when varying the domain length is found using pseudo-arclength continuation previously implemented in Mathematica \cite{Brauns.etal2021}.
The stability of the stationary patterns is determined by linearizing the discretized equations and calculating the eigenvalues of the resulting Jacobian matrix of the linearized dynamics.
This analysis is implemented in Mathematica 13.1.

Setup files for the Comsol simulations, the Python scripts, and the Mathematica notebooks used for the analysis are available at \url{https://github.com/dmuramat/chemotaxis_induced_phase_separation}.